\documentclass[sigconf,screen,nonacm]{acmart}

\usepackage[utf8]{inputenc}
\usepackage[T1]{fontenc}

\usepackage{amsmath}
\usepackage{amssymb}

\usepackage[frozencache, cachedir=.]{minted}

\usepackage{graphicx}
\usepackage{tikz}
\usetikzlibrary{positioning,shapes.geometric,arrows.meta,fit,calc}

\usepackage{pgfplots}
\pgfplotsset{compat=1.18}

\usepackage{booktabs}
\usepackage{array}
\usepackage{colortbl}
\usepackage{makecell}
\usepackage{siunitx}
\usepackage{xcolor}
\definecolor{lightgray95}{RGB}{242,242,242} % ~95%
\definecolor{lightgray90}{RGB}{230,230,230}
\definecolor{lightblue90}{RGB}{219,234,254}
\definecolor{lightblue95}{RGB}{235,245,255}
\definecolor{lightgreen90}{RGB}{220,245,225}
\definecolor{lightgreen85}{RGB}{205,240,212}
\definecolor{lightpurple90}{RGB}{235,225,250}
\definecolor{dangerRed}{RGB}{200,0,0}
\definecolor{okGreen}{RGB}{0,140,60}
\definecolor{uiblue}{RGB}{30,90,180}

\usepackage[most]{tcolorbox}

\tcbset{
  myblock/.style n args={3}{
    enhanced,
    colback=#1,
    colframe=#2,
    boxrule=#3,
    arc=2pt,
    left=6pt,right=6pt,top=6pt,bottom=6pt,
  },
}

\tikzset{
  box/.style={
    draw,
    rounded corners=2pt,
    inner xsep=6pt, inner ysep=6pt,
    align=center
  },
  smallnote/.style={font=\scriptsize},
  boldnote/.style={font=\bfseries\scriptsize},
}

\usepackage{xurl}

\hypersetup{
  colorlinks=true,
  linkcolor=blue,
  filecolor=magenta,
  urlcolor=blue,
  citecolor=blue,
  breaklinks=true
}
\renewcommand\footnotetextcopyrightpermission[1]{} % suppress rights footnote

\title{SecureSign: Bridging Security and UX in Mobile Web3 through Emulated EIP-6963 Sandboxing}

\author{Charles Cheng Ji}
\affiliation{%
  \institution{Refract Labs}
  \city{New York}
  \state{New York}
  \country{USA}
}
\affiliation{%
  \vspace{0.3em} % 
  \institution{New York University}
  \city{New York}
  \state{New York}
  \country{USA}
}
\email{cheng@refract.network}

\author{Brandon Kong}
\affiliation{%
  \institution{Refract Labs}
  \city{New York}
  \state{New York}
  \country{USA}
}
\affiliation{%
  \vspace{0.3em} % 
  \institution{University of Pennsylvania}
  \city{Philadelphia}
  \state{Pennsylvania}
  \country{USA}
}
\email{brandon@refract.network}

\keywords{Mobile Web3, Wallet Security, Progressive Web Apps, Click-jacking Prevention, dApp Sandboxing, Security Economics, User Adoption Barriers}

\begin{document}

\begin{abstract}
Mobile Web3 faces catastrophic retention ($<5\%$) yielding effective acquisition costs of $\$500-\$1,000$ per retained user. Existing solutions force an impossible tradeoff: embedded wallets achieve moderate usability but suffer inherent click-jacking vulnerabilities; app wallets maintain security at the cost of $2-3\%$ retention due to download friction and context-switching penalties. We present SecureSign, a PWA-based architecture that adapts desktop browser extension security to mobile via EIP-6963 provider sandboxing. SecureSign isolates dApp execution in iframes within a trusted parent application, achieving click-jacking immunity and transaction integrity while enabling native mobile capabilities (push notifications, home screen installation, zero context-switching). Our drop-in SDK requires no codebase changes for existing Web3 applications. Threat model analysis demonstrates immunity to click-jacking, overlay, and skimming attacks while maintaining wallet interoperability across dApps.
\end{abstract}

\maketitle

%%%%%%%%%%%%%%%%%%%%%%%%%%%%%%%%%%%%%%%%%%%%%%%%%%%%%%%%%%%%%%%%%%%%%%%%%%%%%%%%
% BODY
%%%%%%%%%%%%%%%%%%%%%%%%%%%%%%%%%%%%%%%%%%%%%%%%%%%%%%%%%%%%%%%%%%%%%%%%%%%%%%%%

\section{Introduction}

Mobile Web3 is bleeding billions in lost revenue and user value. With more than 50\% of internet traffic coming from mobile devices, the future of decentralized applications (dApps) unquestionably lies in mobile experiences. Yet currently, only 8\% of the top 100 Web3 dApps provide mobile-optimized experiences \cite{binance2024web3}. Those that do face catastrophic retention rates: 75\% of users abandon after day 1, dropping to 12\% by day 5 and just 11\% by day 10 \cite{formo2025cohort}. Six-month retention across Ethereum dApps stands at a dismal 3.55\% \cite{seid2024ethereum}, while retention rates plummet below 5\% immediately following airdrop campaign conclusions \cite{binance2024web3}.

These retention failures occur \textit{after} users have already navigated all the friction points that would be unacceptable in any Web2 mobile application: learning what a wallet is, completing wallet setup, securing seed phrases, understanding gas fees, and authorizing their first blockchain transaction. For an average Web2 user accustomed to instant gratification and one-tap authentication, this onboarding experience represents an insurmountable barrier. The impact is severe: cost per retained user is around $\$500-\$1,000$, confirmed by the $<5\%$ long-term retention rate for airdrop campaigns and $\$42-\$85$ in multiple crypto sectors \cite{binance2024web3}\cite{icoda2024cac}. This is economically unsustainable for most dApp businesses and prohibitive for reaching mainstream audiences.

The financial implications are staggering. Web2 users demonstrate high willingness to pay for seamless mobile experiences, as evidenced by Polymarket's breakthrough mainstream adoption when UX friction was minimized. Yet, the majority of Web3 dApps systematically fail to capture this value, not through lack of product-market fit but through fundamental architectural incompatibilities between mobile platform constraints and Web3 security requirements.

This paper argues that mobile Web3's retention crisis is not an optimization problem amenable to incremental UX improvements, but rather an \textit{architecture problem} requiring fundamental rethinking of how decentralized applications operate on mobile platforms. Current solutions force developers into an impossible tradeoff: sacrifice security for modest usability gains through embedded wallets with inherent click-jacking vulnerabilities, or maintain security through app-based wallets that guarantee user abandonment through download friction and context-switching penalties.

\subsection{Contributions}

We present SecureSign, Refract Labs's solution to the mobile Web3 security-UX dilemma. To summarize, our contributions include:

\begin{enumerate}
    \item \textit{Analysis of the Mobile Web3 Crisis}: We provide a comprehensive breakdown of the user acquisition and retention funnel, quantifying friction points at each stage (discovery, download, wallet setup, first transaction, re-engagement) and explaining why current solutions fail to address fundamental architectural problems.
    
    \item \textit{Threat Model for Existing Solutions}: We present detailed security analyses of embedded wallets, app wallets with dApp browsers, and hybrid "open in new tab" approaches, demonstrating why embedded wallets' iframe-based architecture creates unfixable click-jacking, transaction overlay, and skimming vulnerabilities.
    
    \item \textit{SecureSign Architecture}: We introduce a novel PWA-based sandboxing system that adapts desktop browser extension security models to mobile platforms through EIP-6963 multi-provider discovery \cite{eip6963}, achieving isolation between dApp execution contexts and transaction approval flows without compromising user experience.
    
    \item \textit{Native Mobile Integration}: We demonstrate how PWA architecture enables access to native mobile capabilities—push notifications, home screen installation, offline caching, native share APIs—that traditional Web3 solutions cannot provide, directly addressing retention barriers.
    
    \item \textit{Developer Experience}: We present a drop-in SDK requiring zero codebase changes for existing Web3 applications built on standard tooling (wagmi, RainbowKit, ethers.js), eliminating the costly migration penalties embedded wallet solutions impose.
    
    \item \textit{Formal Threat Model Analysis}: We provide comprehensive security analysis demonstrating SecureSign's immunity to click-jacking, transaction overlay attacks, API manipulation, and credential skimming while maintaining wallet interoperability across dApps.
\end{enumerate}

\subsection{Paper Organization}

Section 2 provides background on Web3 wallet security models, mobile platform constraints, and the EIP-6963 standard. Section 3 analyzes the mobile Web3 crisis through detailed user journey breakdowns with empirical retention data. Section 4 systematically examines current solutions: app wallets, embedded wallets, and hybrid approaches, explaining their security vulnerabilities and UX failures. Section 5 presents SecureSign's architecture, including PWA-based sandboxing, EIP-6963 adaptation, and developer integration. Section 6 details mobile retention optimization through native platform features. Section 7 provides formal threat model analysis and security comparisons. Section 8 discusses mobile-first design tradeoffs. Section 9 covers limitations and future work, and Section 10 concludes.

\section{Background and Related Work}

\subsection{Web3 Wallet Security Models}

Web3 applications require users to cryptographically sign transactions using private keys, creating a fundamental security challenge: keys must be accessible for signing operations while remaining protected from malicious actors. Desktop Web3 has converged on the browser extension model \cite{metamask_architecture}, where wallet software runs in a privileged browser extension context isolated from webpage content through the browser's security sandbox \cite{sandbox_security}. Webpages cannot access extension memory, making private key theft impossible even if a user visits a malicious dApp.

The extension injects a JavaScript provider object into webpages, exposing only safe methods like transaction signing requests. When a user attempts a transaction, the extension displays an approval UI in its own context—completely outside the webpage's control—showing transaction details the webpage cannot manipulate. After user approval, the extension signs the transaction and returns only the signature to the webpage, never exposing the private key.

This model provides strong security guarantees: \textit{visual integrity} (users see authentic transaction details), \textit{transaction integrity} (webpages cannot modify displayed information), and \textit{key isolation} (private keys never enter webpage context). However, mobile browsers do not support extensions, creating a fundamental incompatibility requiring architectural innovation rather than incremental adaptation.

Embedded wallets attempt to solve mobile access through Multi-Party Computation (MPC) \cite{mpc_wallets}, distributing key shares across user devices and vendor servers. While MPC provides cryptographic sophistication, Section 4.2 demonstrates why rendering wallet interfaces in iframes creates unfixable security vulnerabilities regardless of underlying key management.

\subsection{Mobile Platform Constraints}

Mobile operating systems impose security restrictions that fundamentally differ from desktop environments. Mobile browsers cannot install extensions. Context switching between applications—routine on desktop—imposes severe penalties on mobile \cite{mobile_context_switching}: users must exit the current app, navigate the home screen or app switcher, open another app, complete an action, and return to the original app, assuming they remember to do so. Each context switch introduces $10-15\%$ abandonment risk \cite{formo2025retention}.

Progressive Web Apps (PWAs) \cite{push_notification_engagement} emerged as a middle ground between native apps and browser-based experiences. PWAs are web applications that can be installed to the device home screen, run in full-screen mode without browser chrome, access native capabilities like push notifications and offline storage, and receive automatic updates without app store approval. Critically for Web3 applications, PWAs can leverage service workers to create client-side sandboxing architectures not possible in standard mobile browsers.

\subsection{EIP-6963: Multi Injected Provider Discovery}

The Ethereum Improvement Proposal 6963 \cite{eip6963} standardizes how multiple wallet providers can coexist in the same browser context without conflicts. Before EIP-6963, browser extensions raced to inject a global \texttt{window.ethereum} object, causing conflicts when multiple wallets were installed. EIP-6963 introduces an event-based discovery mechanism: extensions announce their presence via \texttt{eip6963:announceProvider} events containing provider metadata (name, icon, UUID) and an EIP-1193-compatible provider object. DApps discover available wallets by listening for these events and allowing users to select their preferred provider.

This standard enables SecureSign's architecture: by implementing EIP-6963 within a PWA sandboxing context, we replicate browser extension security properties on mobile platforms where extensions cannot exist.

\section{The Mobile Web3 Crisis}\label{sec:crisis}

Mobile Web3's user acquisition and retention crisis stems from an architecture that forces users through friction points unacceptable in modern mobile applications. This section analyzes each stage of the user journey, quantifying drop-off rates and explaining why current architectures guarantee abandonment.

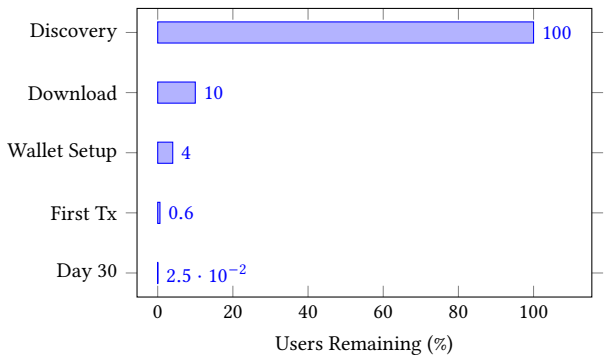
\begin{figure}[t]
\centering
\begin{tikzpicture}
\begin{axis}[
    xbar,
    width=0.9\columnwidth,
    height=5cm,
    font=\small,
    xlabel={Users Remaining (\%)},
    symbolic y coords={Day 30, First Tx, 
                       Wallet Setup, Download, Discovery},
    ytick=data,
    nodes near coords,
    nodes near coords align={horizontal},
    xmin=0, xmax=110,
    bar width=8pt,
    enlarge x limits=0.05,  % Add padding
    y=0.8cm,  % Control y-spacing
]
\addplot coordinates {
    (0.025,Day 30) 
    (0.6,First Tx) 
    (4,Wallet Setup) 
    (10,Download) 
    (100,Discovery)
};
\end{axis}
\end{tikzpicture}
\caption{Mobile Web3 user journey showing catastrophic drop-off}
\label{fig:funnel}
\end{figure}

\subsection{Stage 0: The Opportunity and the Barrier}

The addressable market for mobile Web3 is massive. Web2 mobile users demonstrate high willingness to pay for quality experiences—mobile gaming generates $100+$ billion annually \cite{coinlaw2024demographics}, and mobile fintech apps achieve $40-50\%$ user retention at 90 days. Polymarket's 2024 breakthrough into mainstream consciousness demonstrated that when UX friction is minimized, Web2 users will engage with Web3 applications.

Yet Web3 systematically fails to capture this market. The barrier is not lack of demand but architectural incompatibility. Current Web3 mobile dApps must either:
\begin{itemize}
    \item Implement custom embedded wallets, sacrificing interoperability and forcing users to re-onboard for each new dApp
    \item Direct users to install third-party wallet apps, requiring 100-200MB downloads and seed phrase backup before the user can even experience the dApp
    \item Accept that their application will remain desktop-only, abandoning $50\%+$ of potential users
\end{itemize}

All three options result in prohibitively high effective CAC ($400-1,000$ per retained user, derived from typical $20-50$ mobile CAC divided by $<5\%$ long-term retention) and retention rates below 5\% at 6 months \cite{seid2024ethereum}.

\subsection{Stage 1: Discovery to Download}

Mobile dApps lack viable distribution channels. Consider a user discovering a dApp through a TikTok advertisement, a Telegram link, or a friend's recommendation. On desktop, clicking the link immediately loads the dApp in the browser, where MetaMask or another extension provides seamless wallet connectivity. The user experiences the product within seconds.

On mobile, this flow breaks entirely. The dApp developer faces an impossible choice:

\paragraph{Option 1: Embedded Wallet} Implement a custom wallet directly in the dApp, allowing immediate access but creating a security nightmare (detailed in Section 4.2). Additionally, this wallet only works for \textit{this} dApp—existing Web3 users must create new wallets and transfer assets, while new users cannot carry their wallet to other dApps. This violates Web3's core principle of portable identity.

\paragraph{Option 2: App Wallet Requirement} Direct users to install a mobile wallet app (Phantom, Rainbow, MetaMask Mobile). The user must:
\begin{itemize}
    \item Exit the dApp they just discovered
    \item Navigate to the app store
    \item Download a 100-200MB application
    \item Open the wallet app
    \item Complete setup (covered in Stage 2)
    \item Remember to return to the original dApp
    \item Find the dApp's URL again
\end{itemize}

This flow guarantees massive abandonment. The dApp's marketing spend effectively advertises for third-party wallet apps rather than driving users to the product. Conversion rates from ad click to wallet installation typically fall below 10\% \cite{formo2025retention}.

\paragraph{Option 3: Give Up on Mobile} Many dApp developers simply abandon mobile support, accepting that they will only reach desktop users—effectively forfeiting $50\%+$ of potential users.

The result: only 8\% of top Web3 dApps have mobile versions \cite{binance2024web3}.

\subsection{Stage 2: The Wallet Setup Cliff}

Users who successfully navigate Stage 1 encounter the wallet setup process—the primary retention barrier in Web3 onboarding. Creating a non-custodial wallet requires:

\begin{enumerate}
    \item \textit{Understanding Conceptual Complexity}: Users must grasp that they are creating a cryptographic identity, not just an account. There is no password reset or customer support that can recover access.
    
    \item \textit{Seed Phrase Backup}: Users must write down a 12-24 word recovery phrase, often their first exposure to cryptographic security practices. Many users take screenshots (insecure), skip the backup entirely (guaranteeing future loss), or fail to understand the importance.
    
    \item \textit{Security Verification}: Many wallets require users to re-enter selected seed phrase words to confirm backup, adding friction.
    
    \item \textit{Network and Token Selection}: Users must understand blockchain networks (Ethereum, Solana, etc.) and often must acquire native tokens for gas fees before using the dApp.
\end{enumerate}

Each step introduces abandonment risk. Studies of mobile onboarding flows demonstrate that each additional required field reduces completion rates by $5-20\%$ \cite{web_performance_impact}. Wallet setup requires dozens of steps.

For dApps using third-party wallet apps, the problem is exacerbated by context switching. The user must:
\begin{itemize}
    \item Exit the dApp
    \item Open the wallet app (if they remember which one they were told to download)
    \item Complete setup in the wallet app's UI, which the dApp developer does not control and cannot optimize
    \item Return to the dApp (assuming the mobile browser didn't reload, losing state)
\end{itemize}

Mobile context switching studies show that each app switch reduces task completion probability by $10-15\%$ \cite{mobile_context_switching}. The wallet setup flow requires at least two context switches, often more.

\textit{Impact}: Industry data shows approximately $60-70\%$ abandonment during wallet setup for users who reached this stage \cite{formo2025retention}.

\subsection{Stage 3: First Transaction}

Users completing wallet setup face their first transaction. Unlike Web2 apps where user actions execute immediately, Web3 requires:

\begin{enumerate}
    \item \textit{Gas Acquisition}: Users must own native blockchain tokens (ETH, SOL, etc.) to pay transaction fees. For new users, this requires:
    \begin{itemize}
        \item Understanding what gas is and why it's required
        \item Finding an on-ramp (exchange or credit card purchase)
        \item Completing KYC verification (often multi-day process)
        \item Purchasing tokens
        \item Waiting for transfer to wallet
    \end{itemize}
    
    \item \textit{Transaction Approval}: The wallet displays a transaction prompt filled with technical information: contract addresses, gas limits, gas prices, hexadecimal data. For complex transactions, users see nested function calls and parameter encodings incomprehensible to non-developers.
    
    \item \textit{Transaction Wait Time}: After approval, users wait for blockchain confirmation—anywhere from seconds to minutes depending on network congestion and gas payment.
\end{enumerate}

This flow violates every principle of modern mobile UX design, which prioritizes instant gratification and zero-friction actions. Where TikTok, Instagram, and mobile games provide immediate feedback to every user action, Web3 demands that users understand gas markets, approve cryptographic operations, and wait for distributed consensus.

\textit{Impact}: Only 15\% of users who connect wallets to dApps complete a first transaction \cite{binance2024web3}, meaning approximately 85\% abandon after connecting but before becoming "active users."

\subsection{Stage 4: Re-engagement Failure}

Users who complete their first transaction are not retained users—they must return repeatedly to generate value for the dApp and themselves. This is where mobile Web3 faces its most insurmountable barrier.

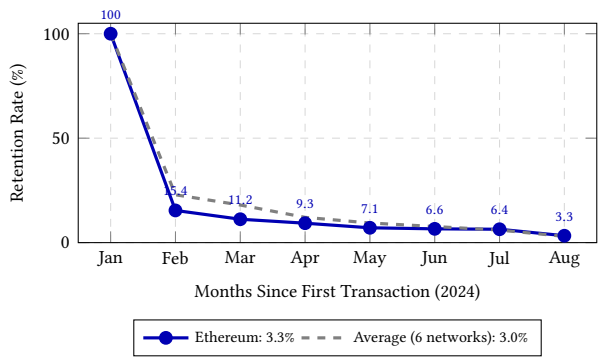
\begin{figure}[t]
\centering
\begin{tikzpicture}
\begin{axis}[
    width=\columnwidth,
    height=4.5cm,
    font=\footnotesize,
    xlabel={Months Since First Transaction (2024)},
    ylabel={Retention Rate (\%)},
    xmin=0.5, xmax=8.5,
    ymin=0, ymax=105,
    xtick={1,2,3,4,5,6,7,8},
    xticklabels={Jan,Feb,Mar,Apr,May,Jun,Jul,Aug},
    grid=major,
    grid style={dashed, gray!30},
    legend style={
        at={(0.5,-0.35)},
        anchor=north,
        legend columns=2,
        font=\scriptsize,
    },
]

% Ethereum (highlighted with values on points)
\addplot[
    color=blue!70!black, 
    mark=*, 
    very thick,
    nodes near coords,
    every node near coord/.append style={
        font=\tiny,
        anchor=south,
        yshift=2pt
    }
] coordinates {
    (1,100) (2,15.4) (3,11.2) (4,9.3) (5,7.1) (6,6.6) (7,6.4) (8,3.3)
};
\addlegendentry{Ethereum: 3.3\%}

% Average of 6 other networks (BNB, Arbitrum, Avalanche, OP, Starknet, Manta)
\addplot[color=gray, very thick, dashed] coordinates {
    (1,100) (2,22.9) (3,18.0) (4,12.1) (5,9.4) (6,7.7) (7,6.0) (8,3.0)
};
\addlegendentry{Average (6 networks): 3.0\%}

\end{axis}
\end{tikzpicture}
\caption{Mobile Web3 retention decay comparing Ethereum against the average of six other networks, showing 3.0--3.3\% retention after 7 months\protect\footnotemark\cite{binance2024web3}}
\label{fig:retention-networks}
\end{figure}
\footnotetext{Analysis includes Ethereum, BNB Chain, Arbitrum, Avalanche, OP Mainnet, Starknet, and Manta. TON and Base excluded due to anomalous positive retention changes (TON: 32.2\%→35.2\% May-Jun; Base: 30.6\%→34.1\% Mar-Apr) inconsistent with expected monotonic decay patterns.}

\subsubsection{Bookmarks Don't Work on Mobile}

Desktop Web3 relies heavily on browser bookmarks. Users bookmark their favorite dApps and access them with a single click from the constantly visible bookmark bar. Mobile browsers \textit{hide} bookmark bars by default to maximize screen space. Accessing a bookmarked site on mobile requires:
\begin{itemize}
    \item Opening the browser app
    \item Tapping the address bar
    \item Tapping the bookmarks icon (typically a small target)
    \item Scrolling through an unorganized list of bookmarks
    \item Tapping the desired dApp
\end{itemize}

This 5-step process takes 15-30 seconds and requires the user to remember the dApp's name. Compared to opening a native app (single tap on home screen icon, $<1$ second), bookmarks impose massive friction.

\subsubsection{No Re-engagement Triggers}

Mobile app retention depends heavily on push notifications and home screen presence \cite{push_notification_engagement}. Apps send notifications reminding users to return ("Your daily reward is ready!", "John just beat your high score"). The app icon provides constant visual reminder on the home screen.

Browser-based dApps cannot send push notifications. Users must remember to check back manually. Without notification triggers, retention drops precipitously: studies show that mobile apps with push notifications achieve $8\times$ higher engagement than those without \cite{push_notification_engagement}.

\subsubsection{Context Switching Remains Expensive}

Even if users remember to return, they face the same context-switching costs:
\begin{itemize}
    \item For app wallet users: Open browser $\rightarrow$ find dApp $\rightarrow$ wait for load $\rightarrow$ perform action $\rightarrow$ triggers wallet approval $\rightarrow$ context switch to wallet app $\rightarrow$ approve $\rightarrow$ context switch back to dApp $\rightarrow$ wait for confirmation
    \item For embedded wallet users: Open browser $\rightarrow$ find dApp $\rightarrow$ perform action $\rightarrow$ approve (no context switch, but security compromised per Section 4.2)
\end{itemize}

Each context switch adds 10-15 seconds and introduces abandonment risk \cite{mobile_context_switching}.

\subsubsection{The Retention Freefall}

The compounding effect of these barriers creates retention freefall:
\begin{itemize}
    \item Day 1-2: User completes first transaction, interested and engaged
    \item Day 3-4: User forgets to return; no push notification reminder
    \item Day 5-7: User attempts to return but cannot find bookmark or remember URL
    \item Day 8-10: User tries context-switching flow, finds it clunky, abandons mid-transaction
    \item Day 30: User joins the 97\% who have churned \cite{seid2024ethereum}
\end{itemize}

\textit{Impact}: 75\% abandon after Day 1, 88\% by Day 5, 89\% by Day 10 \cite{formo2025retention}. Six-month retention: 3.55\% \cite{seid2024ethereum}.

% \begin{figure}[t]
% \centering
% \setlength{\tabcolsep}{8pt}
% \renewcommand{\arraystretch}{1.2}
% \begin{tabular}{@{} >{\raggedright\arraybackslash}p{0.55\linewidth} >{\raggedleft\arraybackslash}p{0.3\linewidth} @{}}
% \toprule
% \textbf{Metric} & \textbf{Cost per Retained User} \\
% \midrule
% Typical Web2 Mobile CAC & \$20--50 \\
% Web3 Effective CAC ($\div$ 0.05 retention) & \$400--1{,}000 \\
% \midrule
% \textbf{Penalty Multiplier} & \textbf{20$\times$} \\
% \bottomrule
% \end{tabular}
% \caption{The retention penalty multiplier effect on customer acquisition costs}
% \label{fig:cac-penalty}
% \end{figure}

\subsection{The Impossible Tradeoff}

These retention failures stem from fundamental architectural constraints, not poor UX design. Mobile Web3 faces an impossible tradeoff: \textit{maintain security through app-based wallets and accept 2-3\% retention, or sacrifice security through embedded wallets to achieve moderate usability}. Section 4 demonstrates why every current approach fails to resolve this tradeoff.

\section{Analysis of Current Solutions}\label{sec:current}

This section analyzes three dominant approaches to mobile Web3 wallet integration, demonstrating why each fails to simultaneously achieve security and retention.

\subsection{App Wallets with dApp Browsers}

Mobile wallet applications like Phantom, Rainbow, and MetaMask Mobile embed web browsers within the wallet app, allowing users to access dApps without leaving the wallet's security context. Private keys remain in the wallet app's secure storage, and the embedded browser provides a JavaScript provider for transaction signing.

\subsubsection{Architecture}

The wallet app packages a mobile browser engine (WebView on Android, WKWebView on iOS) configured to inject a wallet provider into loaded webpages. When the user navigates to a dApp within the wallet's browser, the page receives a \texttt{window.ethereum} or similar provider object. Transaction requests flow from the webpage to the wallet app, which displays approval UI in the native app context. After approval, the wallet signs the transaction and returns the signature to the webpage.

This architecture provides strong security isolation: the dApp runs in a browser context with no access to the wallet app's memory or storage, and approval UI renders in the native app layer where webpages cannot manipulate it.

\subsubsection{The Customer Acquisition Cost Problem}

App wallets require users to discover, download, and set up a standalone application before they can access any dApp. As detailed in Section 3.2, this creates prohibitive acquisition costs:

\begin{itemize}
    \item \textit{Download Friction}: 100-200MB app downloads on mobile data connections discourage installation. Users discovering a dApp through social media or advertisements must abandon their discovery context, navigate to app stores, wait for downloads, and then return—each step introducing abandonment risk.
    
    \item \textit{Market Education Cost}: Marketing spend promoting a dApp effectively advertises for third-party wallet providers. Users learn "I need Phantom to use Web3" rather than remembering the specific dApp they wanted to try.
    
    \item \textit{Limited Market}: App wallet requirements restrict dApp audiences to Web3-native users who have already installed wallets.
\end{itemize}

\subsubsection{The Re-engagement Failure}

Even after successful onboarding, app wallets create retention barriers:

\begin{itemize}
    \item \textit{Navigation Friction}: Users must open the wallet app, navigate to the browser tab, access their browsing history or bookmarks (within the wallet app's browser), and find the desired dApp. This 4-5 step process competes poorly against one-tap native app opening.
    
    \item \textit{Unoptimized for dApp Use}: Wallet apps primarily optimize for trading, staking, and portfolio management. The embedded browser is a secondary feature, often with limited functionality and slow performance compared to dedicated browsers.
    
    \item \textit{No Persistent Presence}: Users see the wallet app icon on their home screen, not individual dApp icons. There is no direct path from "I want to play that game" to opening the game—users must remember that the game lives inside their wallet app's browser.
    
    \item \textit{Limited Notification Support}: While wallet apps can send push notifications, individual dApps cannot trigger notifications about dApp-specific events. A gaming dApp cannot notify users about daily rewards; a DeFi dApp cannot alert users to liquidation risks.
\end{itemize}

\subsubsection{Retention Impact}

App wallet approaches achieve $2-3\%$ retention at 30 days \cite{formo2025retention}, slightly worse than already-abysmal Web3 averages. The security model is sound, but user experience guarantees abandonment.

\subsection{Embedded Wallets}

Embedded wallet solutions like Privy, Magic, and Dynamic render wallet functionality directly within dApp webpages using iframes and server-side key management. Users authenticate via email, social login, or SMS, and the embedded wallet provider manages private keys on the user's behalf, typically using Multi-Party Computation (MPC) to distribute key shares \cite{mpc_wallets}.

\subsubsection{Architecture}

The dApp embeds an iframe hosted on the wallet provider's domain. When a user signs up, they authenticate through this iframe, and the wallet provider's servers generate key shares—one stored on the user's device (in browser storage), one on the provider's servers. Transaction signing requires coordinating both shares via MPC protocols.

Transaction approval UI renders within the iframe. From the dApp's perspective, the wallet is simply another UI component. Users never leave the dApp webpage, eliminating context switching.

\subsubsection{The Promise: Elimination of Download Friction}

Embedded wallets remove app download requirements. Users sign up with email or social login, familiar flows requiring seconds rather than minutes. No seed phrase backup, no app store navigation, no 200MB downloads. For dApp developers, embedded wallets promise SDK-based integration and control over the onboarding flow.

\subsubsection{The Fatal Flaw: Unfixable Security Vulnerabilities}

Despite usability improvements, embedded wallets suffer from fundamental security vulnerabilities inherent to iframe-based architectures. These are not implementation bugs fixable through better coding practices—they are architectural limitations that cannot be resolved while maintaining the embedded model \cite{hackernews2024iframe}.

\paragraph{Click-Jacking and Transaction Overlay Attacks}
\textit{Threat Model}: A malicious dApp embeds an iframe containing a legitimate embedded wallet. The iframe displays a transaction approval prompt: "Approve 5 USDC to purchase NFT." However, the parent dApp webpage uses CSS positioning to overlay a pixel-perfect fake transaction prompt on top of the real iframe.

The fake prompt shows "Approve 10 USDC purchase" while the real underlying iframe shows "Transfer 1000 USDC to 0xAttackerAddress." Because iframes are rendered as rectangular regions within the parent document, the parent can position absolutely-positioned DOM elements on top, making the iframe invisible while showing convincing fake UI \cite{clickjacking2008}.

\begin{figure*}[t]
\centering
\includegraphics[width=0.9\textwidth]{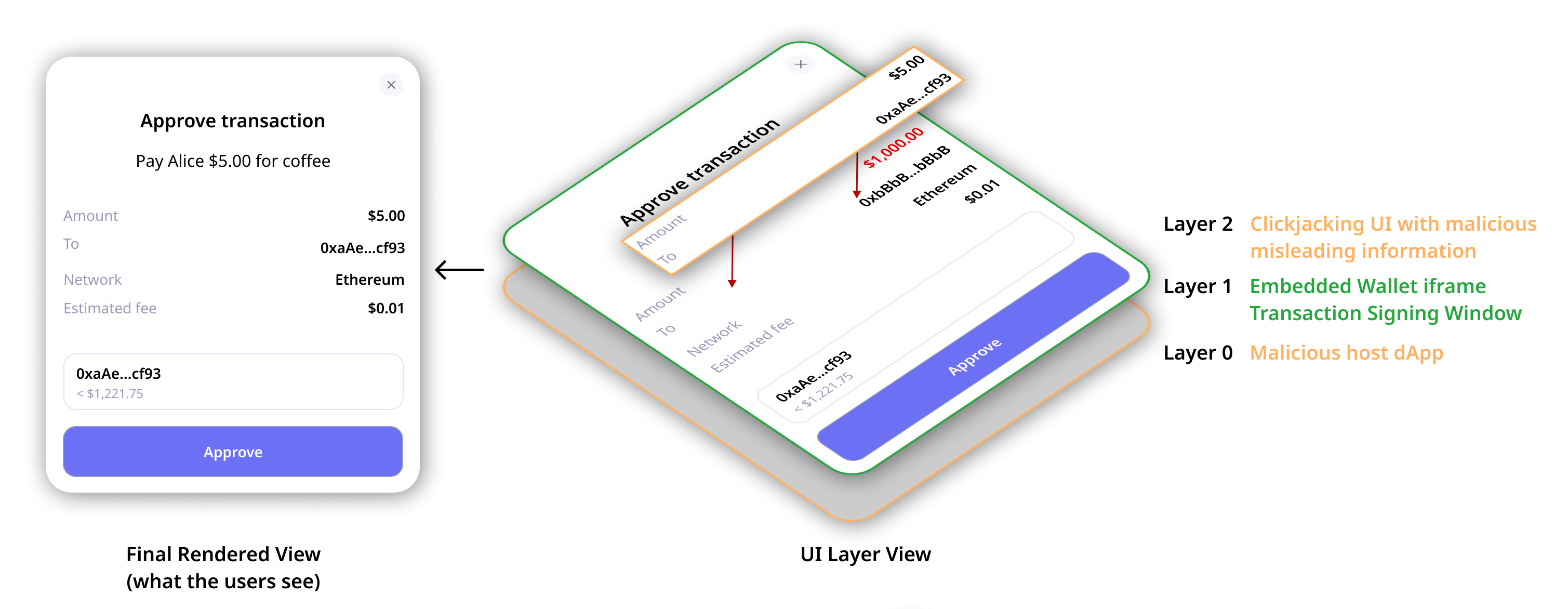}
\caption{Click-jacking attack on embedded wallets: malicious dApp overlays fake UI on top of legitimate iframe approval prompt, allowing transaction manipulation (yellow highlighted region)}
\label{fig:clickjacking}
\end{figure*}

When the user clicks "Approve" on the fake overlay, the click event passes through to the iframe underneath, clicking the real approval button for the malicious transaction.

\textit{Why It's Unfixable}: The fundamental issue is that \textit{the parent document controls the visual rendering context}. The browser composites the iframe's content into the parent document's rendering tree, allowing the parent to manipulate what the user actually sees. The iframe itself has no mechanism to detect overlays or prevent parent manipulation.

Some embedded wallet providers attempt mitigations:
\begin{itemize}
    \item Frame-busting scripts that detect when the iframe is hidden (opacity, z-index, position)
    \item Requiring the iframe to be visible and sufficiently large
    \item Monitoring for suspicious CSS properties
\end{itemize}

However, these are cat-and-mouse games. Attackers can:
\begin{itemize}
    \item Use timing attacks: make overlays visible only during the approval interaction
    \item Exploit browser quirks and rendering edge cases
    \item Use sophisticated CSS and SVG techniques to evade detection \cite{clickjacking2008}
\end{itemize}

\paragraph{Automated Click-Jacking via JavaScript}
Even if overlays are detected, malicious dApps can programmatically trigger clicks on iframe elements without user action:

\begin{minted}[
    breaklines,
    fontsize=\footnotesize,
    breakanywhere=true
]{tsx}
const iframe = document.getElementById('embedded-wallet');
const approveButton = iframe.contentWindow.document.querySelector('.approve-btn');
approveButton.click(); // Simulated click
\end{minted}

While same-origin policy prevents direct DOM manipulation of cross-origin iframes, browsers allow dispatching synthetic events that can interact with iframe content in limited ways. Combined with social engineering (e.g., "Click anywhere to continue"), attackers can trigger unintended approvals.

\paragraph{Credential and Payment Skimming}
Embedded wallets with fiat on-ramps allow users to purchase cryptocurrency with credit cards, rendering payment forms within iframes. Malicious dApps can overlay fake payment forms that capture credit card details before submitting them to the legitimate payment processor:

\textit{Attack}: Dapp overlays pixel-perfect fake credit card form on top of embedded wallet's real form. User enters credit card number, CVV, billing address into the fake form. The fake form captures the data (sending it to the attacker) and then programmatically fills the real iframe form underneath, completing the transaction so the user doesn't suspect compromise. The attacker now has full credit card details for fraud \cite{hackernews2024iframe}.

\begin{figure*}[t]
\centering
\includegraphics[width=0.9\textwidth]{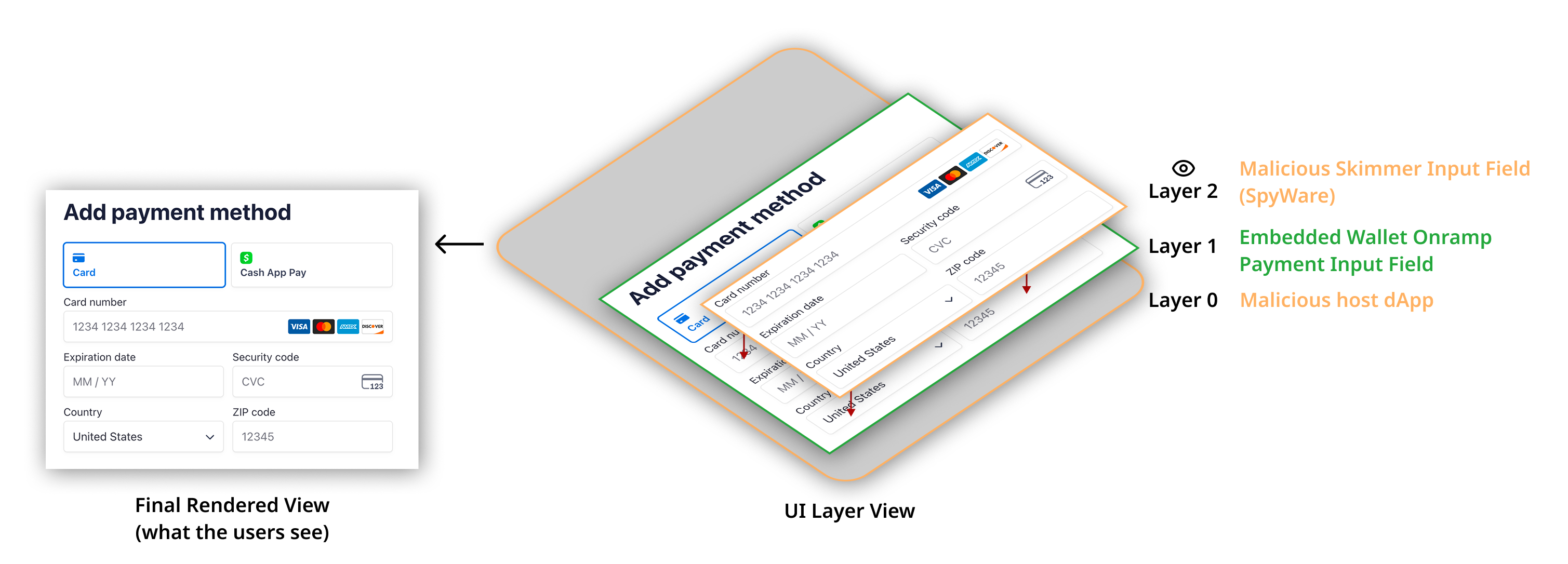}
\caption{UI skimming attack: malicious dApp overlays fake payment form (yellow highlighted region) to capture credit card details before submitting to legitimate processor}
\label{fig:skimming}
\end{figure*}

This attack succeeded against 49 merchants using Stripe payment iframes in 2024 \cite{hackernews2024iframe}, demonstrating that even sophisticated payment processors cannot fully protect iframe-based forms.

\paragraph{Why Embedded Wallets Cannot Provide Security Indicators}
Desktop browsers show security indicators for HTTPS connections: padlock icons, domain names in the address bar, EV certificate organization names. Users learn to check these indicators before entering sensitive information.

Iframes provide \textit{no equivalent security indicators visible to users}. The parent webpage's address bar shows the dApp's domain, not the wallet provider's domain. Users have no way to verify that the transaction approval UI they see is genuine and not an attacker overlay. Browser extensions and native apps can provide verified UI with security indicators; iframes fundamentally cannot.

\subsubsection{The Interoperability Sacrifice}

Because of these security vulnerabilities, embedded wallet providers make a critical design choice: \textit{each dApp receives an isolated wallet}. Even if a user signs into two different dApps with the same email address, they receive two different wallet addresses. This prevents a compromised dApp from draining wallets used in other dApps.

However, this breaks a core Web3 principle: \textit{portable user-owned identity}. Users cannot carry assets between dApps, must on-ramp separately for each dApp (re-entering credit cards multiple times), and lose the network effects of consistent on-chain identity. This is unintuitive and terrible UX—users expect one wallet, not dozens.

\subsubsection{The Developer Penalty}

Embedded wallets require substantial codebase changes. Web3 applications built on industry-standard libraries (wagmi, RainbowKit, Web3Modal, ethers.js) use client-side signing patterns: the application requests a signature from a browser-injected provider, which returns a signed message. Embedded wallets replace this with server-side APIs: the application sends the transaction to the wallet provider's servers, which coordinate MPC signing and return the signature.

This architectural shift requires:
\begin{itemize}
    \item Replacing wallet connection logic throughout the codebase
    \item Migrating transaction signing calls from client-side libraries to vendor APIs
    \item Handling authentication flows and session management differently
    \item Rewriting state management for wallet interactions
    \item Testing edge cases where users want traditional wallet connections
\end{itemize}

For teams with existing production Web3 applications, migration costs range from hundreds to thousands of engineering hours. For early-stage projects, this technical debt diverts resources from feature development with no guarantee of retention improvement.

\subsubsection{Vendor Lock-In}

Embedded wallet pricing scales with monthly active wallets, typically $0.05-0.15$ per MAW. At scale, this becomes a significant operational expense: 100,000 MAW costs $5,000-15,000$/month, 1,000,000 MAW costs $50,000-150,000$/month.

More critically, migration is nearly impossible. Users' private keys exist only in the vendor's infrastructure. Switching providers requires:
\begin{itemize}
    \item Every user creating a new wallet with the new provider
    \item Every user manually transferring all assets to the new wallet
    \item Abandoning wallet addresses with associated on-chain history and reputation
    \item Re-onboarding the entire user base
\end{itemize}

This makes vendor switching so costly that it's effectively impossible once a dApp has substantial users, giving vendors pricing power and creating single points of failure. If the vendor experiences outages, regulatory issues, or business failure, the dApp's entire user base is impacted with no fallback.

\subsubsection{Summary: Modest UX Gains, Catastrophic Security Loss}

Embedded wallets improve onboarding conversion from $\sim$10\% (app wallets) to $\sim$60\% but achieve this by sacrificing security through unfixable architectural vulnerabilities, destroying interoperability through wallet isolation, and locking developers into proprietary vendors. This is not an acceptable tradeoff for applications managing user funds.

\subsection{"Open in New Tab" Approaches}

A third approach, exemplified by Para wallet, attempts to split the difference: wallet operations open in new browser tabs rather than embedded iframes, avoiding iframe security issues while eliminating app downloads.

\subsubsection{Architecture}

When a user connects a wallet or signs a transaction, the dApp opens the wallet provider's URL in a new browser tab via \texttt{window.open()}. The wallet tab renders approval UI, obtains user consent, and communicates the result back to the original dApp tab via \texttt{postMessage} or URL parameters.

\subsubsection{Why It Fails: Browser Popup Blocking}

Mobile browsers, particularly Safari (iOS's only browser engine), \textit{aggressively block popup windows}. Safari blocks \texttt{window.open()} calls by default, showing a small, easily-missed notification that a popup was blocked. Users clicking "Connect Wallet" see nothing happen. They click again, confused. Maybe a tiny notification appears. They must:
\begin{itemize}
    \item Notice the popup was blocked (many users don't)
    \item Navigate to Safari settings
    \item Find the popup blocking setting
    \item Allow popups for that specific domain
    \item Return to the dApp
    \item Try connecting again
\end{itemize}

This flow is worse than app wallet downloads—at least app downloads show clear progress. Popup blocking appears broken, leading users to assume the dApp is non-functional.

\subsubsection{Multi-Tab Confusion}

Even when popups work, mobile browsers handle multiple tabs poorly:
\begin{itemize}
    \item Tabs are not visually obvious like desktop browser tabs
    \item Switching between tabs requires 3-4 taps through the tab switcher
    \item Users often don't realize a new tab opened
    \item Return navigation is non-intuitive—users must manually switch back to the original tab
\end{itemize}

\subsubsection{Security Issues Remain}

Opening in a new tab doesn't solve security problems—it merely shifts them. The wallet creation and authentication flows still occur in a standard browser tab, providing no verified security indicators. Attackers can:
\begin{itemize}
    \item Register typo-squatted domains (\texttt{walletnprovider.com} vs \texttt{walletprovider.com})
    \item Create pixel-perfect phishing pages
    \item Inject malicious code via compromised browser extensions (if the user has any)
\end{itemize}

Users have no way to verify that the new tab is legitimate, especially on mobile where address bars are tiny and often hidden.

\subsubsection{Limited Adoption}

The "open in new tab" approach has seen minimal adoption, with Para wallet being the primary example. The approach fails on all dimensions: Safari blocks it, UX is worse than alternatives, and security remains compromised. It represents the worst of both worlds.

\section{SecureSign Architecture}\label{sec:architecture}

SecureSign resolves the mobile Web3 security-UX incompatibility by adapting the proven desktop browser extension security model to mobile platforms through Progressive Web App (PWA) based sandboxing. This section details our architecture, explaining how we achieve desktop-class security isolation while providing superior mobile UX compared to app wallets and embedded wallets.

\subsection{Design Principles}

SecureSign's architecture rests on three principles:

\begin{enumerate}
    \item \textit{No Compromise on Security}: Transaction approval UI must render in a context that dApps cannot manipulate, providing the same isolation guarantees as desktop browser extensions. Users must see verifiable security indicators distinguishing legitimate approval prompts from phishing attempts.
    
    \item \textit{No Compromise on UX}: Mobile users expect one-tap access, instant load times, push notifications, and seamless flows without context switching. We leverage native mobile platform capabilities rather than fighting against platform constraints.
    
    \item \textit{Zero Migration Cost for Developers}: Existing Web3 applications built on industry-standard tooling (wagmi, RainbowKit, ethers.js, Web3Modal) must integrate SecureSign with minimal code changes, avoiding the migration penalty embedded wallets impose.
\end{enumerate}

\subsection{Overview: PWA-Based Sandboxing}

SecureSign consists of two components:

\begin{itemize}
    \item \textit{Refract Passport}: A Progressive Web App serving as the parent application and trusted security context. Passport manages wallet operations, renders transaction approval UI, and provides access to native mobile capabilities. Passport cannot be manipulated by individual dApps.
    
    \item \textit{Refract SDK}: A lightweight JavaScript library that dApps embed. The SDK provides an EIP-6963-compliant wallet provider that integrates seamlessly with existing Web3 tooling, requiring no changes to existing wallet connection or transaction signing code.
\end{itemize}

When a dApp is accessed through Refract Passport, the dApp runs in a sandboxed iframe within the Passport PWA. The SDK establishes a secure \texttt{postMessage} communication channel between the iframe and the parent Passport application, replicating the browser extension security model where webpages cannot directly access extension memory or UI.

\subsection{Detailed Architecture}

\subsubsection{Component 1: Refract Passport PWA}

Refract Passport is installed as a PWA on the user's mobile device, providing:

\begin{itemize}
    \item \textit{Wallet Key Management}: User private keys are managed via Multi-Party Computation (MPC) \cite{mpc_wallets} with key shares distributed between the device's secure storage (leveraging browser IndexedDB with encryption) and Refract's backend infrastructure. Signing operations require coordination between both shares, preventing single-point compromise.
    
    \item \textit{dApp Isolation via iframes}: Each dApp accessed through Passport loads in a sandboxed iframe with a unique origin. The browser's same-origin policy prevents iframes from accessing parent application memory, DOM, or storage. Critically, \textit{Passport controls the iframe rendering and cannot be manipulated by iframe content}, inverting the trust model from embedded wallets where the parent dApp controls the iframe wallet.
    
    \item \textit{Transaction Approval UI}: All transaction approval prompts render in the parent Passport application context, completely outside any iframe (shown in Figure~\ref{fig:securesign-clickjack}). dApps send transaction requests via \texttt{postMessage}, but Passport's approval UI displays transaction details parsed independently. The dApp cannot overlay, hide, or manipulate this UI because it renders in the parent context where iframes have no control \cite{sandbox_security}.

\begin{figure*}[t]
\centering
\includegraphics[width=0.9\textwidth]{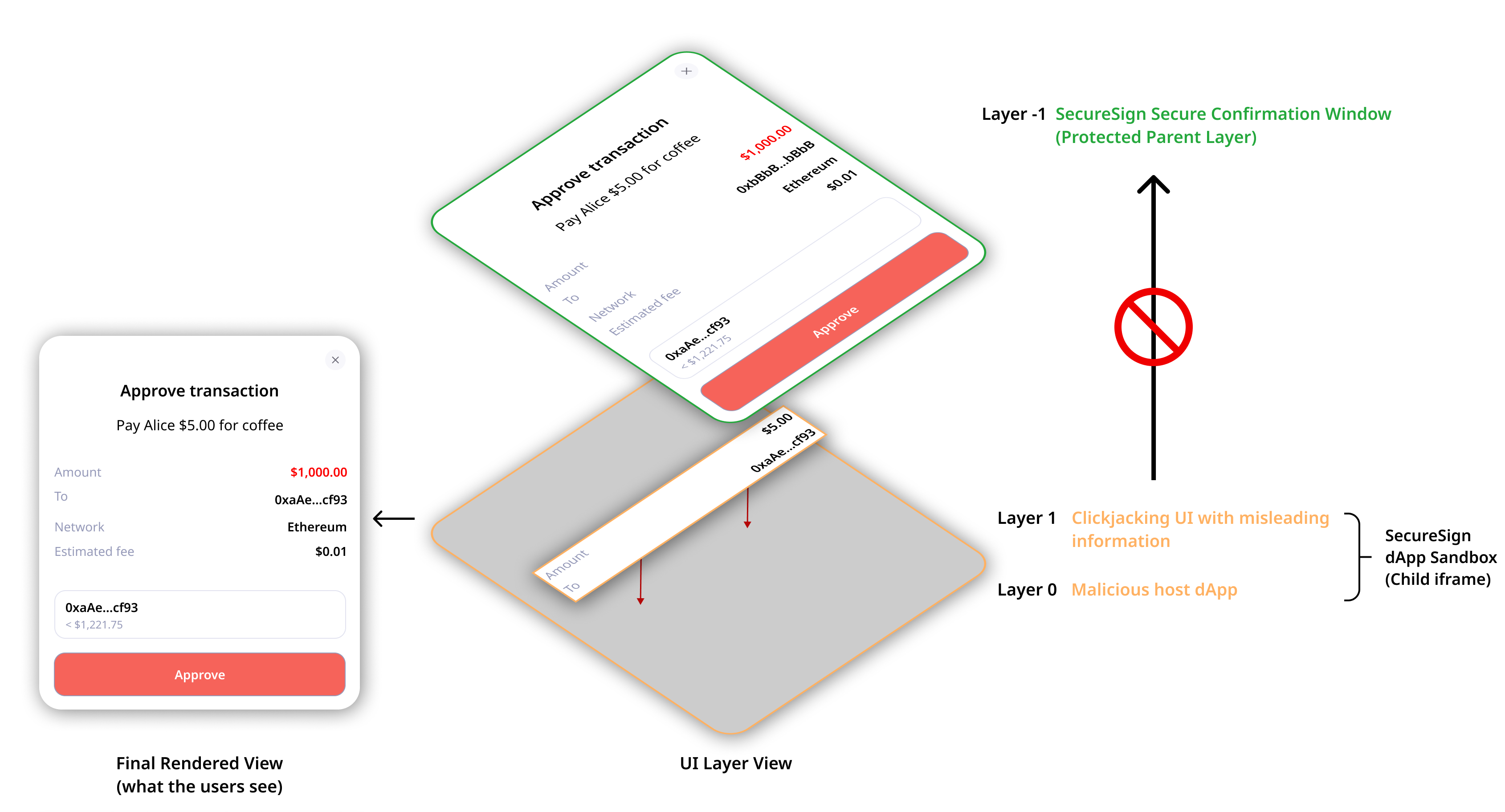}
\caption{SecureSign's click-jacking protection: approval UI renders in parent Passport context, isolated from dApp iframe manipulation}
\label{fig:securesign-clickjack}
\end{figure*}
    
    \item \textit{Native Mobile Capabilities}: As a PWA installed to the home screen, Passport can access mobile platform APIs unavailable to browser tabs:
    \begin{itemize}
        \item Push notifications via Service Worker Push API
        \item Home screen icon presence providing persistent visual reminder
        \item Full-screen rendering without browser chrome
        \item Offline caching via Service Workers
        \item Background sync for pending transactions
        \item Native share sheet integration
    \end{itemize}
    
    \item \textit{dApp Registry and Verification}: Passport maintains a registry of verified dApps with associated metadata (name, logo, domain, appId). When displaying approval UI, Passport shows verified dApp identity, preventing phishing attacks where malicious sites impersonate legitimate dApps.
\end{itemize}

\subsubsection{Component 2: Refract SDK and EIP-6963 Provider}

The Refract SDK provides a drop-in provider that integrates with existing Web3 tooling:

\begin{minted}[
    breaklines,
    fontsize=\footnotesize,
    breakanywhere=true
]{tsx}
import { InjectProvider } from "@refract-network/inject";

function App() {
  return (
    <InjectProvider
      appId="your-dapp-id"
      parentOrigin="https://passport.refract.network">
      {/* Existing Web3 setup - no changes needed */}
      <WagmiConfig config={wagmiConfig}>
        <RainbowKitProvider>
          {/* Your app components */}
        </RainbowKitProvider>
      </WagmiConfig>
    </InjectProvider>
  );
}
\end{minted}

The \texttt{InjectProvider} wraps the existing application, detecting whether the dApp is running inside Refract Passport. If so, it injects an EIP-6963-compliant provider:

\begin{itemize}
    \item \textit{Provider Injection}: The SDK dispatches \texttt{eip6963:\allowbreak announce\allowbreak Provider} events containing Refract's provider metadata and an EIP-1193-compatible provider object. Web3 libraries like wagmi automatically discover this provider via EIP-6963, treating it as any other wallet \cite{eip6963}.
    
    \item \textit{PostMessage Communication Bridge}: The provider object exposes standard methods (\texttt{eth\_\allowbreak requestAccounts}, \texttt{eth\_\allowbreak sendTransaction}, \texttt{personal\_\allowbreak sign}, etc.). When called, these methods serialize the request and send it to the parent Passport application via \texttt{window.\allowbreak parent.\allowbreak postMessage()}, including the authenticated \texttt{appId} for authorization.
    
    \item \textit{Response Handling}: Passport processes the request, displays approval UI to the user, performs signing if approved, and sends the result back to the iframe via \texttt{postMessage}. The SDK's provider receives the response and resolves the original promise, making the entire process transparent to the dApp code.
    
    \item \textit{Graceful Fallback}: If the dApp is accessed in a standard browser (not via Passport), the SDK detects the absence of the parent Passport context and does not inject its provider. Standard browser extension wallets (MetaMask, etc.) remain available via EIP-6963, allowing the dApp to function on desktop without modification.
\end{itemize}

\subsubsection{Security Properties}

This architecture provides several critical security properties:

\begin{figure*}[t]
\centering
\includegraphics[width=0.7\textwidth]{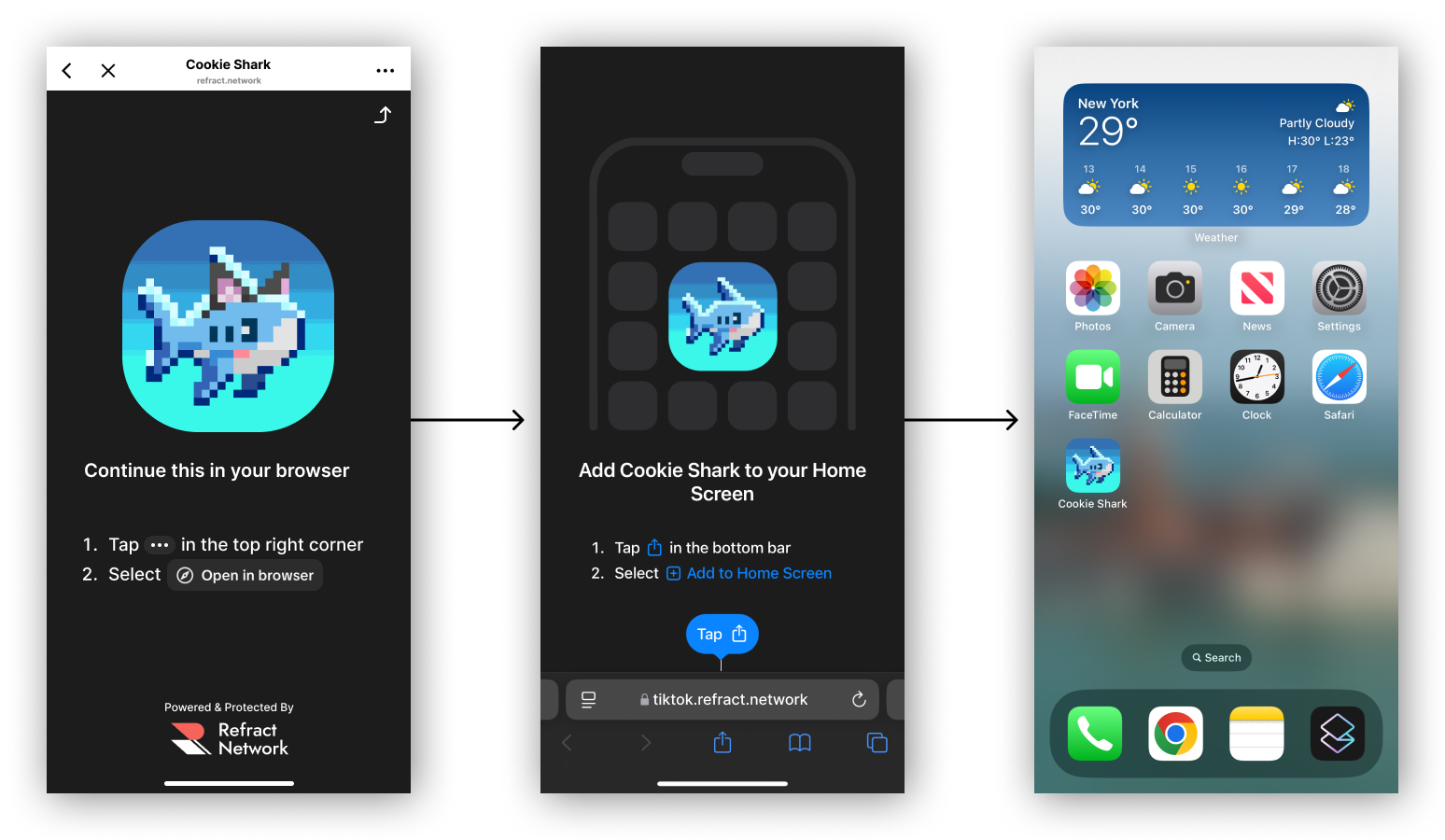}
\caption{PWA installation flow: users add dApps to home screen for one-tap access, eliminating bookmark navigation friction}
\label{fig:homescreen}
\end{figure*}

\begin{itemize}
    \item \textit{Isolation}: dApps run in iframes and cannot access Passport's memory, storage, or DOM. The browser's same-origin policy enforces this isolation at the platform level \cite{sandbox_security}.
    
    \item \textit{Visual Integrity}: Transaction approval UI renders in Passport's context, not the dApp's iframe. The iframe cannot overlay, hide, or manipulate approval prompts. Users see transaction details that Passport independently parses, not details the dApp controls.
    
    \item \textit{Click-Jacking Immunity}: Because approval UI is in the parent application, click-jacking attacks are impossible. The dApp cannot create overlays on Passport's UI because the rendering contexts are isolated by iframe boundaries.
    
    \item \textit{Transaction Integrity}: Passport receives transaction requests via \texttt{postMessage}, parses them independently using trusted libraries, and displays parsed details to users. Even if a dApp sends malicious transaction data, Passport shows the true transaction details, allowing users to reject suspicious transactions.
    
    \item \textit{API Authentication}: Every transaction request includes the dApp's registered \texttt{appId}. Passport verifies that the requesting origin matches the registered dApp domain, preventing unauthorized transaction requests from malicious sites attempting to impersonate legitimate dApps.
    
    \item \textit{Phishing Prevention}: Passport's approval UI displays verified dApp identity (name, logo, domain) pulled from Refract's dApp registry, not from dApp-provided data. Users can verify they're interacting with the intended application, and attackers cannot impersonate verified dApps without compromising Refract's registry infrastructure.
\end{itemize}

\section{Mobile Retention Optimization}\label{sec:retention}

Security alone does not solve retention. SecureSign's PWA architecture enables access to native mobile capabilities that directly address the retention barriers identified in Section 3.4, providing features that traditional Web3 solutions fundamentally cannot offer.

\subsection{PWA Installation: Solving the Bookmark Problem}

Mobile browsers hide bookmark bars, making bookmarks a poor retention mechanism (Section 3.4). PWAs solve this through home screen installation.

\subsubsection{Implementation}

When a user first accesses a dApp through Refract Passport, the SDK can trigger an installation prompt:

\begin{minted}[
    breaklines,
    fontsize=\footnotesize,
    breakanywhere=true
]{tsx}
const { requestPWAInstall, requestPWAStatus } = useInjectProvider();

// Check if already installed
const status = await requestPWAStatus();
if (!status.installed) {
  // Prompt installation at appropriate moment
  await requestPWAInstall();
}
\end{minted}

The browser displays a native installation prompt: "Add CookieShark to Home Screen" (shown in Figure~\ref{fig:homescreen}). Upon acceptance, an icon appears on the device home screen providing one-tap access.

\subsubsection{User Experience}

Installed dApps launch with a single tap from the home screen, identical to native apps. The PWA opens in full-screen mode without browser chrome (no address bar, no browser UI), providing an immersive experience indistinguishable from native apps. The home screen icon provides persistent visual presence, reminding users the dApp exists.

\subsubsection{Retention Impact}

Studies show PWA users demonstrate $2-3\times$ higher retention rates than browser-only users \cite{push_notification_engagement}. The constant visual reminder on the home screen combined with one-tap access dramatically reduces the friction of returning to the dApp.

\subsection{Push Notifications: Solving Re-engagement}

Browser-based dApps cannot send push notifications, eliminating a primary mobile retention mechanism (Section 3.4). Installed PWAs access the Push API, enabling notifications that appear on the device lock screen.

\subsubsection{Implementation}

dApps request notification permission via the Refract SDK:

\begin{minted}[
    breaklines,
    fontsize=\footnotesize,
    breakanywhere=true
]{tsx}
const { sendNotification } = useInjectProvider();

// Request permission
const permission = await requestNotificationPermission();

if (permission === 'granted') {
  // Send notification from backend
  await sendNotification({
    title: "Daily Reward Available!",
    message: "Claim your 50 tokens now",
    interaction: true, // Tapping opens the dApp
    icon: "/icon.png",
    badge: "/badge.png"
  });
}
\end{minted}

Notifications are sent via Refract's backend to ensure delivery even when the PWA is not open. Tapping a notification opens the dApp directly, providing seamless re-engagement.

\subsubsection{Use Cases}

\begin{itemize}
    \item \textit{Transaction Confirmations}: Notify users when transactions complete ("Your NFT purchase succeeded!")
    \item \textit{Time-Sensitive Events}: Alert users to limited-time opportunities ("Airdrop ending in 1 hour")
    \item \textit{Social Interactions}: Notify about social activity ("Alice just challenged you to a game")
    \item \textit{Reward Reminders}: Encourage daily engagement ("Don't miss your login streak bonus!")
\end{itemize}

\subsubsection{Retention Impact}

Push notification engagement rates in mobile apps range from $40-50\%$ opt-in \cite{push_notification_engagement}, and users with notifications enabled show $8\times$ higher engagement rates than those without. Notifications provide the critical re-engagement trigger that browser-based dApps lack.

\subsection{Zero Context Switching: Integrated Approval Flow}

App wallet approaches require context switching between the dApp and wallet app for every transaction, imposing 10-15 second delays and abandonment risk (Section 3.4). SecureSign's integrated architecture eliminates context switching entirely.

\subsubsection{Implementation}

When a dApp requests transaction approval, Passport's approval UI slides up as a modal overlay on top of the dApp content. The dApp remains visible underneath (with dimmed overlay), and the approval UI contains:
\begin{itemize}
    \item Verified dApp identity (name, logo, domain)
    \item Human-readable transaction summary parsed from transaction data
    \item Gas fee estimate
    \item Approve/Reject buttons
\end{itemize}

After approval, the modal dismisses and the dApp continues immediately, with no app switching or navigation.

\subsubsection{User Experience}

The user never leaves the dApp context. The approval flow feels like a standard mobile modal dialog—familiar, fast, and intuitive. The dApp state is preserved throughout approval, eliminating reload risks.

Swipe gestures work naturally: swiping down dismisses the approval modal (rejecting the transaction), matching iOS modal conventions. The back button also dismisses the modal, providing multiple interaction affordances.

\subsubsection{Performance Impact}

Context switching adds 10-15 seconds per transaction (Section 3.4). SecureSign's integrated approval flow requires $<2$ seconds from approval prompt to completion, an $80\%+$ time reduction. Each eliminated context switch reduces abandonment risk by $10-15\%$ \cite{mobile_context_switching}.

\subsection{Native Share API: Viral Growth Enablement}

Web2 mobile apps leverage native share functionality for viral growth: users tap "Share" and the OS presents a native share sheet with installed apps (Messages, WhatsApp, Twitter, etc.). Web dApps have limited access to this functionality; PWAs unlock full native share integration.

\subsubsection{Implementation}

\begin{minted}[
    breaklines,
    fontsize=\footnotesize,
    breakanywhere=true
]{tsx}
const { requestShare } = useInjectProvider();

await requestShare({
  title: "Join me on CookieShark!",
  text: "Play and earn tokens together",
  url: "https://cookieshark.refract.network?ref=user123",
  // Automatically handled: deep linking back to PWA
});
\end{minted}

This triggers the OS native share sheet, allowing users to share via any installed app. Refract automatically embeds referral tracking codes and ensures shared links open in Refract Passport with the referring user's code captured.

\subsubsection{Viral Growth Mechanisms}

\begin{itemize}
    \item \textit{Referral Incentives}: dApps can reward users for successful referrals, creating viral growth loops
    \item \textit{Social Proof}: Shared messages include user-specific context ("Alice achieved level 10!")
    \item \textit{Frictionless Sharing}: One-tap sharing reduces completion time from 30+ seconds (manual URL copy/paste) to $<5$ seconds
\end{itemize}

\subsubsection{Impact}

Native share APIs increase share completion rates by 60\% compared to manual URL copying \cite{push_notification_engagement}. Each retained user bringing 1-3 additional users through referrals amplifies customer acquisition efficiency.

\subsection{Service Worker Caching: Instant Loading}

Mobile networks are often slow and unreliable. Standard web dApps reload all assets on every visit, creating 3-5 second blank screens that drive immediate abandonment—47\% of users abandon sites that take $>3$ seconds to load \cite{web_performance_impact}.

\subsubsection{Implementation}

SecureSign's PWA architecture includes Service Worker caching:

\begin{minted}[
    breaklines,
    fontsize=\footnotesize,
    breakanywhere=true
]{tsx}
// Automatic in Refract SDK
self.addEventListener('install', (event) => {
  event.waitUntil(
    caches.open('dapp-assets-v1').then((cache) => {
      return cache.addAll([
        '/index.html',
        '/app.js',
        '/styles.css',
        '/logo.png',
        // All static assets
      ]);
    })
  );
});

self.addEventListener('fetch', (event) => {
  event.respondWith(
    caches.match(event.request).then((response) => {
      // Return cached version, fetch update in background
      return response || fetch(event.request);
    })
  );
});
\end{minted}

On first visit, assets are cached. Subsequent visits load from cache in $<100$ms while background-fetching updated content. The user sees content immediately, and updates apply after load.

\subsubsection{Performance Impact}

\begin{itemize}
    \item \textit{First Load}: 2-3 seconds (standard web performance)
    \item \textit{Subsequent Loads}: $<100$ms ($90\%+$ faster)
    \item \textit{Offline Functionality}: dApps continue working without network connectivity, showing cached content and queuing transactions for later submission
    \item \textit{Data Usage Reduction}: 90\% reduction in data transfer after first visit, important for users on limited data plans
\end{itemize}

\subsubsection{Retention Correlation}

Load time directly correlates with retention: every additional second of load time reduces conversion by 7\% \cite{web_performance_impact}. SecureSign's near-instant loading on repeat visits removes a major abandonment driver.

\subsection{Comparative Retention Infrastructure}

The following table summarizes native capabilities available to each approach:

\begin{figure*}[t]
\centering
\setlength{\tabcolsep}{8pt}
\renewcommand{\arraystretch}{1.2}
\rowcolors{2}{gray!10}{white}
\begin{tabular}{@{} >{\raggedright\arraybackslash}p{0.35\textwidth}
                  >{\centering\arraybackslash}p{0.15\textwidth}
                  >{\centering\arraybackslash}p{0.18\textwidth}
                  >{\centering\arraybackslash}p{0.18\textwidth} @{}}
\toprule
\textbf{Feature} & \textbf{App Wallets} & \textbf{Embedded} & \textbf{SecureSign} \\
\midrule
Home Screen Installation & \checkmark\ (wallet app) & \texttimes & \checkmark\ (per dApp) \\
Push Notifications & \checkmark\ (wallet only) & \texttimes & \checkmark\ (per dApp) \\
Zero Context Switching & \texttimes & \checkmark & \checkmark \\
Native Share API & \texttimes & \texttimes & \checkmark \\
Offline Caching & Limited & \texttimes & \checkmark \\
One-Tap Access & \texttimes & Bookmark & \checkmark \\
Full-Screen Mode & \checkmark & \texttimes & \checkmark \\
\bottomrule
\end{tabular}
\caption{Comparison of native mobile capabilities across solutions}
\label{fig:capabilities}
\end{figure*}

SecureSign is the only solution providing comprehensive native mobile integration. App wallets cannot provide per-dApp home screen icons or notifications because dApps run in an embedded browser within the wallet app. Embedded wallets lack these features because they're iframe-based within standard browsers that don't support PWA capabilities for embedded content.

\section{Threat Model Analysis}\label{sec:threat}

This section provides formal security analysis of SecureSign's threat model, examining potential attack vectors and demonstrating our security properties.

\subsection{Attack Surface}

We consider three attacker profiles:

\begin{itemize}
    \item \textit{Malicious dApp Developer}: A developer creates a dApp that appears legitimate but attempts to steal user funds or credentials. The dApp is accessed through Refract Passport, giving it the ability to send transaction requests and display UI within its iframe.
    
    \item \textit{Compromised dApp}: A previously legitimate dApp is compromised through supply chain attacks, XSS vulnerabilities, or server breaches. Attackers inject malicious code into the dApp's frontend.
    
    \item \textit{Network Attacker}: An attacker with network-level access (malicious WiFi, ISP-level, nation-state) attempts man-in-the-middle attacks, phishing via DNS spoofing, or traffic manipulation.
\end{itemize}

We assume Refract Passport's infrastructure and the mobile device OS remain uncompromised. Attacks compromising the Passport PWA itself or device root/jailbreak are out of scope, as these undermine all security models.

\subsection{Security Properties and Defenses}

\subsubsection{Transaction Approval Integrity}

\begin{itemize}
    \item \textit{Property}: Users see authentic transaction details that the dApp cannot manipulate.
    
    \item \textit{Threat}: Malicious dApp attempts transaction overlay attack: displays "Approve 10 USDC payment" while submitting transaction transferring 1000 USDC to attacker address.
    
    \item \textit{Defense}: Transaction approval UI renders in Refract Passport's parent context, outside the dApp's iframe. The dApp sends transaction data via \texttt{postMessage}, but Passport independently parses this data using trusted libraries and displays parsed human-readable details. The dApp's UI cannot overlay or manipulate Passport's approval modal because iframe content cannot affect parent context rendering \cite{sandbox_security}.
    
    \item \textit{Attack Attempt}:
\end{itemize}
\begin{minted}[
    breaklines,
    fontsize=\footnotesize,
    breakanywhere=true
]{tsx}
// In dApp iframe
const fakeUI = document.createElement('div');
fakeUI.innerHTML = 'Approve 10 USDC payment';
fakeUI.style.cssText = 'position: fixed; z-index: 9999;';
document.body.appendChild(fakeUI);

// This overlay only affects the iframe's context, not Passport's
// Passport's approval UI renders in parent context, unaffected
\end{minted}
\begin{itemize}
    \item \textit{Result}: Attack fails. The user sees Passport's approval UI showing the true transaction details (1000 USDC transfer), allowing them to reject the malicious transaction.
\end{itemize}

\subsubsection{Click-Jacking Immunity}

\begin{itemize}
    \item \textit{Property}: Attackers cannot trigger transaction approvals without explicit user interaction on authentic approval UI.
    
    \item \textit{Threat}: Malicious dApp attempts programmatic clicking on approval buttons or overlays invisible approval UI that users click unintentionally.
    
    \item \textit{Defense}: Approval UI renders in the parent Passport application context. The iframe's JavaScript cannot access the parent context's DOM (prevented by same-origin policy \cite{sandbox_security}), cannot dispatch events to parent context elements, and cannot position overlays outside its iframe boundary.
    
    \item \textit{Attack Attempt}:
\end{itemize}
\begin{minted}[
    breaklines,
    fontsize=\footnotesize,
    breakanywhere=true
]{tsx}
// In dApp iframe
try {
  // Attempt to access parent DOM
  const approveBtn = window.parent.document.querySelector('.approve-button');
  approveBtn.click(); // Attempt programmatic click
} catch (e) {
  // SecurityError: Blocked by same-origin policy
}

// Attempt overlay attack
const overlay = document.createElement('div');
overlay.style.cssText = `
  position: absolute;
  top: 0; left: 0;
  width: 100vw; height: 100vh;
  z-index: 999999;
`;
document.body.appendChild(overlay);
// This overlay is confined to the iframe boundaries
// Cannot extend outside iframe to cover parent UI
\end{minted}
\begin{itemize}
    \item \textit{Result}: Attack fails. The dApp cannot access or manipulate approval UI, and overlays remain confined within iframe boundaries.
\end{itemize}

\subsubsection{API Authentication and Authorization}

\begin{itemize}
    \item \textit{Property}: Only authorized dApps can send transaction requests, and requests are validated against registered dApp metadata.
    
    \item \textit{Threat}: Attacker creates malicious website attempting to impersonate a legitimate dApp, sending transaction requests claiming to be from "TrustedDApp.com".
    
    \item \textit{Defense}: Every transaction request includes the dApp's registered \texttt{appId}. Refract Passport validates:
    \begin{enumerate}
        \item The requesting iframe's origin matches the registered domain for that \texttt{appId}
        \item The \texttt{appId} exists in Refract's dApp registry
        \item The user has authorized this \texttt{appId} to access their wallet
    \end{enumerate}
    If any validation fails, the request is rejected before showing approval UI.
    
    \item \textit{Attack Attempt}:
\end{itemize}

\begin{minted}[
    breaklines,
    fontsize=\footnotesize,
    breakanywhere=true
]{tsx}
// On attacker.com, attempt to impersonate trusted-dapp.com
window.parent.postMessage({
  type: 'eth_sendTransaction',
  appId: 'trusted-dapp-id', // Stolen/guessed appId
  transaction: { to: '0xAttacker', value: '1000000000000000000' }
}, '*');
\end{minted}
\begin{itemize}
    \item \textit{Result}: Passport receives the message, validates the origin (\texttt{attacker.com}) against the registered origin for \texttt{trusted-dapp-id} (\texttt{trusted-dapp.com}), detects mismatch, and rejects the request without showing approval UI. The attack is logged for security monitoring.
\end{itemize}

\subsubsection{Phishing Prevention}

\begin{itemize}
    \item \textit{Property}: Users can verify they are interacting with intended dApps, and attackers cannot impersonate verified dApps.
    
    \item \textit{Threat}: Attacker registers \texttt{trustedapp.com} (vs. legitimate \texttt{trusted-app.com}), creates pixel-perfect clone of legitimate dApp, and tricks users into connecting wallets and approving transactions.
    
    \item \textit{Defense}: Refract Passport's approval UI displays verified dApp identity pulled from the dApp registry, including:
    \begin{itemize}
        \item Official dApp name
        \item Verified logo
        \item Registered domain
        \item Security status (verified checkmark for audited dApps)
    \end{itemize}
    This information comes from Refract's registry, not from the dApp's own code. Attackers cannot modify these indicators without compromising Refract's infrastructure.
    
    Additionally, domain verification requires dApps to prove ownership through DNS records or file-based verification before registration, preventing impersonation.
    
    \item \textit{Attack Attempt}:
\end{itemize}
\begin{minted}[
    breaklines,
    fontsize=\footnotesize,
    breakanywhere=true
]{tsx}
// On trustedapp.com (typosquat), set fake metadata
<meta name="dapp-name" content="Trusted App" />
<meta name="dapp-icon" content="/fake-logo.png" />

// Request transaction
sendTransaction({ to: '0xAttacker', value: '1 ETH' });
\end{minted}
\begin{itemize}
    \item \textit{Result}: Passport's approval UI shows the domain \texttt{trustedapp.com} and name from registry entry (if registered; likely "Unknown dApp" for unverified domains). Users see a mismatch from the expected legitimate domain and reject the transaction. Sophisticated users check the domain and detect the typosquat.
\end{itemize}

\begin{figure*}[t]
\centering
\includegraphics[width=0.9\textwidth]{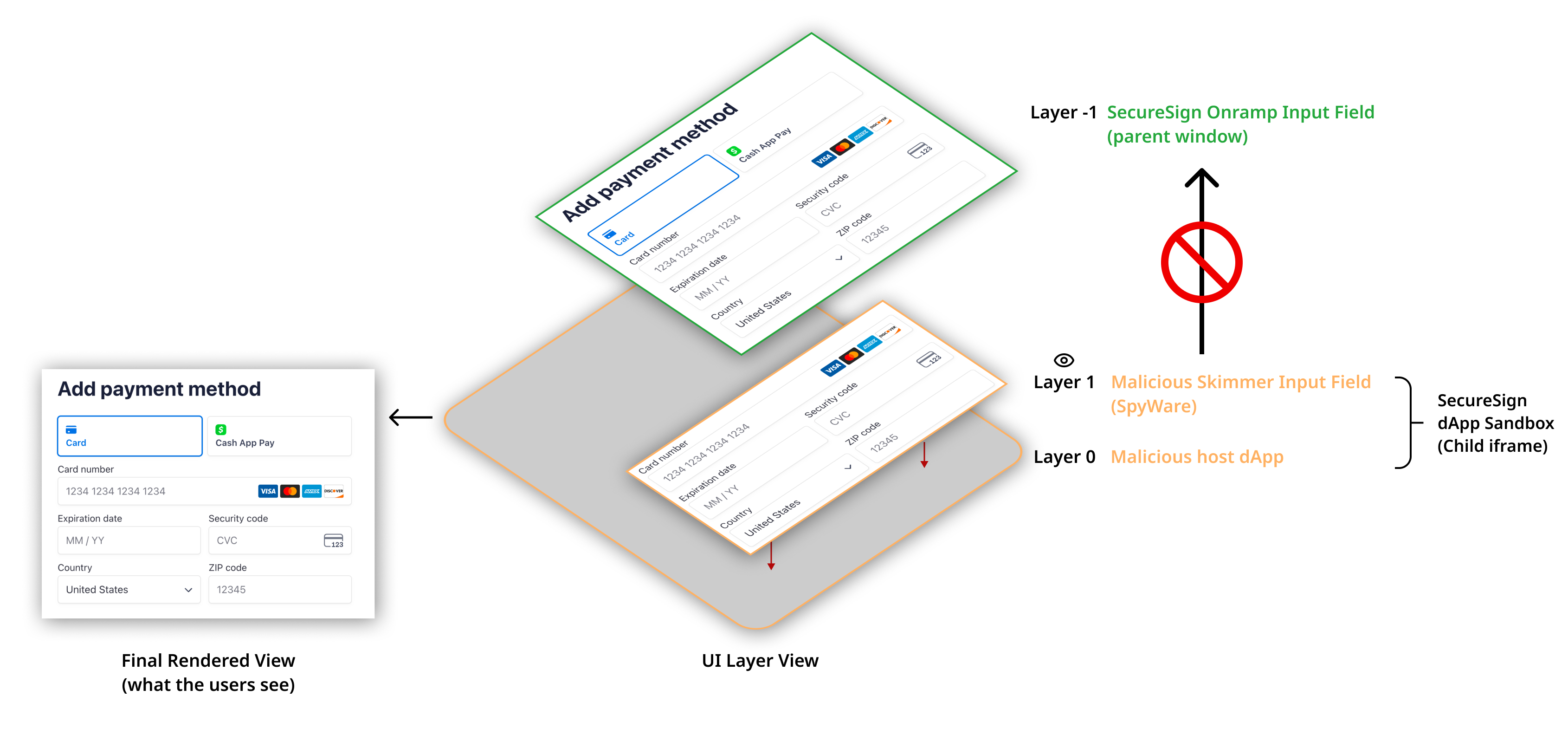}
\caption{SecureSign's credential protection: payment forms render in isolated parent context, preventing dApp access to sensitive data}
\label{fig:securesign-skimming}
\end{figure*}

\subsubsection{Credential Skimming Prevention (Figure~\ref{fig:securesign-skimming})}

\begin{itemize}
    \item \textit{Property}: Credit card and authentication credentials entered during wallet setup or fiat on-ramping cannot be intercepted by dApps.
    
    \item \textit{Threat}: Malicious dApp overlays fake credit card form during fiat on-ramp flow, capturing user's credit card details.
    
    \item \textit{Defense}: Wallet management UI (including on-ramp payment forms) renders in Refract Passport's parent context, not in dApp iframes. Payment forms are hosted on payment processor domains (Stripe, MoonPay) in separate browsing contexts, following PCI-DSS compliant integration patterns. dApps never receive access to payment form contexts.
    
    \item \textit{Attack Attempt}:
\end{itemize}
\begin{minted}[
    breaklines,
    fontsize=\footnotesize,
    breakanywhere=true
]{tsx}
<div style="position: fixed; z-index: 999999;">
  <form action="https://attacker.com/steal">
    <input name="cc" placeholder="Credit Card Number" />
    <input name="cvv" placeholder="CVV" />
    <button>Purchase</button>
  </form>
</div>
\end{minted}
\begin{itemize}
    \item \textit{Result}: The fake form is confined to the dApp's iframe and cannot overlay Passport's payment UI, which renders in a separate context. Users interact with the legitimate payment processor form, and the dApp never receives payment credentials.
\end{itemize}

\subsection{Comparative Vulnerability Analysis}

The following table compares vulnerability exposures across solutions:

\begin{figure*}[t]
\centering
\setlength{\tabcolsep}{8pt}
\renewcommand{\arraystretch}{1.25}
\rowcolors{2}{gray!10}{white}
\begin{tabular}{@{} >{\raggedright\arraybackslash}p{0.40\linewidth}
                  >{\centering\arraybackslash}p{0.15\linewidth}
                  >{\centering\arraybackslash}p{0.20\linewidth}
                  >{\centering\arraybackslash}p{0.20\linewidth} @{}}
\toprule
\textbf{Vulnerability} & \textbf{App Wallets} & \textbf{Embedded Wallets} & \textbf{SecureSign} \\
\midrule
Click-Jacking          & \checkmark\ Immune   & \texttimes\ Vulnerable    & \checkmark\ Immune \\
Transaction Overlay    & \checkmark\ Immune   & \texttimes\ Vulnerable    & \checkmark\ Immune \\
Credential Skimming    & \checkmark\ Immune   & \texttimes\ Vulnerable    & \checkmark\ Immune \\
Phishing (Fake dApp)   & Partial              & Partial                   & \checkmark\ Protected \\
Programmatic Approval  & \checkmark\ Immune   & \texttimes\ Vulnerable    & \checkmark\ Immune \\
API Impersonation      & \checkmark\ N/A      & Possible                  & \checkmark\ Prevented \\
\bottomrule
\end{tabular}
\caption{Security Vulnerability Comparison Across Mobile Web3 Solutions}
\end{figure*}

SecureSign achieves security properties equivalent to app wallets (both based on strong isolation models) while providing superior UX. Embedded wallets remain fundamentally vulnerable to attacks exploiting their iframe architecture.

\subsection{STRIDE Threat Model Analysis}

To provide rigorous security analysis, we apply Microsoft's STRIDE threat modeling framework \cite{stride_threat_modeling}, systematically examining all threat categories across SecureSign's architecture. STRIDE categorizes threats into six classes: Spoofing, Tampering, Repudiation, Information Disclosure, Denial of Service, and Elevation of Privilege.

\subsubsection{Spoofing Identity}

\textbf{Threat:} Adversaries attempt to impersonate legitimate entities (users, dApps, or Passport itself).

\textbf{Attack Vectors:}
\begin{enumerate}
    \item \textit{Malicious dApp Impersonation}: Attacker registers domain similar to legitimate dApp (typosquatting) and attempts to appear as the real dApp
    \item \textit{Passport PWA Phishing}: Attacker creates fake Passport PWA to steal credentials
    \item \textit{Provider Injection Attack}: Malicious code attempts to inject fake EIP-6963 provider before legitimate provider
\end{enumerate}

\textbf{Defenses:}
\begin{itemize}
    \item \textit{dApp Registry with Domain Verification}: Refract maintains registry of verified dApps requiring DNS-based or file-based domain ownership proof before registration. Approval UI displays verified dApp name, logo, and domain from registry (not from dApp-supplied data), making impersonation require compromising Refract's registry infrastructure
    \item \textit{HTTPS Enforcement with Certificate Pinning}: All communication uses HTTPS with certificate validation. Passport PWA served over HTTPS with integrity checks
    \item \textit{EIP-6963 Provider UUID Verification}: Each provider includes cryptographically verifiable UUID. Refract's provider announces with registered UUID that dApps can validate
    \item \textit{Origin Validation}: Every \texttt{postMessage} includes origin header verified against registered dApp domains
\end{itemize}

\textbf{Residual Risks:} Users must verify Passport domain during first installation (trust-on-first-use). DNS hijacking or compromised certificate authorities could enable sophisticated attacks, though these undermine all web-based security models.

\subsubsection{Tampering}

\textbf{Threat:} Adversaries attempt to modify data in transit or storage.

\textbf{Attack Vectors:}
\begin{enumerate}
    \item \textit{Transaction Data Tampering}: Attacker modifies transaction data between dApp request and user approval
    \item \textit{postMessage Interception}: Man-in-the-middle attacks on \texttt{postMessage} communication
    \item \textit{MPC Key Share Tampering}: Attacker attempts to modify stored key shares
    \item \textit{Service Worker Code Injection}: Malicious code injected into Service Worker to modify behavior
\end{enumerate}

\textbf{Defenses:}
\begin{itemize}
    \item \textit{Independent Transaction Parsing}: Passport independently parses transaction data using trusted libraries, displaying parsed details rather than dApp-supplied descriptions
    \item \textit{Message Authentication}: \texttt{postMessage} communication includes \texttt{appId} authentication binding messages to registered origins
    \item \textit{Encrypted Key Share Storage}: Device-side key shares stored in browser IndexedDB with encryption. Server-side shares stored with access controls and audit logging
    \item \textit{Service Worker Integrity}: Service Workers served with Subresource Integrity (SRI) hashes and Content Security Policy (CSP)
    \item \textit{HTTPS Transport Security}: All network communication uses TLS 1.3+ with perfect forward secrecy
\end{itemize}

\textbf{Residual Risks:} Device compromise could allow key share extraction. Backend server compromise remains single point of failure (acknowledged limitation).

\subsubsection{Repudiation}

\textbf{Threat:} Users or dApps deny having performed actions.

\textbf{Defenses:}
\begin{itemize}
    \item \textit{Cryptographic Signing}: All approvals result in cryptographically signed transactions recorded immutably on-chain
    \item \textit{Client-Side Logging}: Passport maintains local logs of approval events
    \item \textit{Backend Audit Trail}: Server-side MPC signing operations logged with timestamps
    \item \textit{Blockchain Finality}: Completed transactions achieve blockchain finality, making repudiation cryptographically infeasible
\end{itemize}

\subsubsection{Information Disclosure}

\textbf{Threat:} Unauthorized access to sensitive information including private keys and transaction history.

\textbf{Defenses:}
\begin{itemize}
    \item \textit{MPC Key Distribution}: Private keys never reconstructed in full. Device and server each hold one share \cite{mpc_wallets}
    \item \textit{Iframe Isolation}: Same-origin policy enforces memory isolation between dApps \cite{sandbox_security}
    \item \textit{TLS Encryption}: All network traffic encrypted via TLS 1.3
    \item \textit{Minimal Data Collection}: Passport collects only essential data
\end{itemize}

\textbf{Residual Risks:} Network metadata visible to observers. Backend has access to one key share and transaction metadata.

\subsubsection{Denial of Service}

\textbf{Defenses:}
\begin{itemize}
    \item \textit{Backend Redundancy}: Deployed across multiple availability zones with load balancing
    \item \textit{Rate Limiting}: Transaction request rate limits per dApp and per user
    \item \textit{User Control}: Users can always reject transactions. Approval prompts timeout after 60 seconds
    \item \textit{Service Worker Fallback}: If Service Worker fails, PWA degrades gracefully
\end{itemize}

\subsubsection{Elevation of Privilege}

\textbf{Defenses:}
\begin{itemize}
    \item \textit{Browser Sandbox}: Same-origin policy enforced at browser engine level \cite{sandbox_security}
    \item \textit{Origin-Based Access Control}: \texttt{postMessage} origin validation ensures only authorized domains send requests
    \item \textit{Registry Access Controls}: Multi-factor authentication and audit logging for registry administration
    \item \textit{Defense in Depth}: Multiple security layers ensure single vulnerability does not compromise entire system
\end{itemize}

\begin{figure*}[t]
\centering
\setlength{\tabcolsep}{8pt}
\renewcommand{\arraystretch}{1.2}
\begin{tabular}{@{} lll @{}}
\toprule
\textbf{Threat Category} & \textbf{Primary Defenses} & \textbf{Risk} \\
\midrule
Spoofing & Registry verification, HTTPS & Low \\
Tampering & Independent parsing, TLS & Low \\
Repudiation & On-chain signatures, logs & Very Low \\
Info Disclosure & MPC, iframe isolation, TLS & Medium \\
Denial of Service & Redundancy, rate limiting & Medium \\
Elevation of Privilege & Browser sandbox, origin ACL & Low \\
\bottomrule
\end{tabular}
\caption{STRIDE Threat Analysis Summary for SecureSign}
\end{figure*}

\subsection{Trust Boundaries and Security Zones}

SecureSign establishes clear trust zones with enforced boundaries:

\begin{itemize}
    \item \textbf{Trusted Zone}: Refract Passport PWA (parent context) --- Transaction Approval UI, MPC Key Share (Device), Service Worker, Registry Verification
    \item \textbf{Trust Boundary}: iframe isolation enforced by browser same-origin policy
    \item \textbf{Untrusted Zone}: dApp iframe (sandboxed) --- dApp Application Code, User-Facing UI, Refract SDK Provider
    \item \textbf{Network Boundary}: TLS 1.3 encryption with certificate pinning
    \item \textbf{Semi-Trusted Zone}: Refract Backend Infrastructure --- MPC Key Share (Server), Push Notification Service, dApp Registry Database
\end{itemize}

\textbf{Key Properties:}
\begin{enumerate}
    \item \textit{Trusted $\rightarrow$ Untrusted}: Information flows controlled. Passport reveals only transaction signatures and account addresses
    \item \textit{Untrusted $\rightarrow$ Trusted}: Flows validated via origin authentication, \texttt{appId} verification, schema validation
    \item \textit{Semi-Trusted Backend}: Backend holds one MPC key share but cannot independently sign transactions
    \item \textit{Network Security}: TLS 1.3 encryption protects all network communication
\end{enumerate}

\subsection{Comparative Security Analysis}

\begin{figure*}[t]
\centering
\setlength{\tabcolsep}{6pt}
\renewcommand{\arraystretch}{1.2}
\rowcolors{2}{gray!5}{white}
\begin{tabular}{@{} >{\raggedright\arraybackslash}p{0.35\linewidth}
                  >{\centering\arraybackslash}p{0.18\linewidth}
                  >{\centering\arraybackslash}p{0.22\linewidth}
                  >{\centering\arraybackslash}p{0.18\linewidth} @{}}
\toprule
\textbf{Security Mechanism} & \textbf{App Wallets} & \textbf{Embedded} & \textbf{SecureSign} \\
\midrule
Approval UI Context & Native app & iframe & PWA parent \\
Transaction Parsing & Trusted code & iframe code & Trusted code \\
Origin Validation & N/A & Limited & Cryptographic \\
Click-Jacking Defense & OS isolation & Detectable & Sandbox \\
Key Storage & OS enclave & MPC & MPC \\
Single Point of Failure & Device loss & Vendor & Backend \\
Cross-dApp Sharing & \checkmark & \texttimes & \checkmark \\
Desktop Compatible & \texttimes & \checkmark & \checkmark \\
Phishing Prevention & App store & Domain bar & Registry \\
Credential Skimming & OS APIs & Vulnerable & Protected \\
\bottomrule
\end{tabular}
\caption{Detailed Security Mechanism Comparison}
\end{figure*}

\subsection{Limitations and Threat Model Boundaries}

\textbf{In-Scope Threats:}
\begin{itemize}
    \item Malicious or compromised dApp developers
    \item Network attackers (MitM, phishing, DNS spoofing)
    \item Cross-dApp attacks
    \item User social engineering (within approval UI design limits)
    \item Browser vulnerabilities under active exploit
\end{itemize}

\textbf{Out-of-Scope Threats (acknowledged limitations):}
\begin{itemize}
    \item \textit{Refract Backend Compromise}: Server-side key share compromise could enable DoS. \textit{Future mitigation}: decentralized backend or threshold signing with multiple servers
    \item \textit{Device Compromise}: Malware with root access can extract device key share (affects all mobile wallets)
    \item \textit{Browser Zero-Days}: Undisclosed vulnerabilities could violate isolation (affects all web security models)
    \item \textit{Physical Device Access}: Attacker with unlocked device can approve transactions
    \item \textit{Clipboard Attacks}: Malware modifying clipboard to replace wallet addresses
\end{itemize}

\textbf{Trust Assumptions:}
\begin{enumerate}
    \item Browser implements same-origin policy correctly
    \item TLS certificate authorities not compromised for Refract domains
    \item User device OS provides basic application isolation
    \item Refract backend infrastructure secured against intrusion
    \item User can recognize Passport PWA domain during initial installation
\end{enumerate}

These assumptions align with standard web security models and mobile application security practices.

\section{Design Trade-offs and Mobile-First Decisions}\label{sec:tradeoffs}

SecureSign makes deliberate architectural choices optimized for mobile Web3. Some perceived limitations are intentional design decisions that maximize security and UX within mobile platform constraints.

\subsection{Magic dApp Link Requirement}

Each dApp requires a unique, persistent URL ("Magic dApp Link") serving as its primary entry point (e.g., \url{https://cookieshark.refract.network}).

\subsubsection{Rationale}

\begin{itemize}
    \item \textit{Application Identity Verification}: Binding URLs to registered \texttt{appId} values enables domain verification and phishing prevention. Passport validates that iframe origins match registered dApp domains.
    
    \item \textit{PWA Installation Requirements}: PWAs require stable URLs for home screen installation. Changing URLs invalidate installations.
    
    \item \textit{Deep Linking}: Push notifications and social shares use URLs to open specific dApps. Persistent URLs enable reliable deep linking.
    
    \item \textit{Security Boundary Enforcement}: Unique origins provide natural iframe security boundaries enforced by same-origin policy \cite{sandbox_security}.
\end{itemize}

\subsubsection{Why This Works on Mobile}

\begin{itemize}
    \item \textit{Telegram Mini Apps}: Already use persistent bot links (\url{https://t.me/botname/app}). Users familiar with this model.
    
    \item \textit{PWA URL Hiding}: Installed PWAs hide the URL bar after initial load. Users never see URLs during normal usage.
    
    \item \textit{Social Share Automation}: Native share API copies URLs automatically. Users don't manually handle URLs.
    
    \item \textit{Deep Linking Infrastructure}: Mobile OSs have built-in deep linking (\url{https://} links can open installed PWAs directly).
\end{itemize}

\subsubsection{Developer Impact}

One-time setup during dApp registration. Deploy dApp at designated URL. Zero runtime overhead. Better SEO from persistent URLs compared to dynamically generated paths.

\subsection{PWA Architecture Dependency}

SecureSign requires Progressive Web App technology, not native iOS/Android apps.

\subsubsection{Rationale}

\begin{itemize}
    \item \textit{Iframe Sandboxing with Security Policies}: PWAs can configure iframe sandboxing policies that standard web pages cannot. Service Workers enable enforcing security policies at the network layer.
    
    \item \textit{Native Mobile Capabilities}: PWAs access push notifications, home screen installation, offline caching, and native share APIs without app store distribution.
    
    \item \textit{Cross-Platform Single Codebase}: One codebase works on iOS, Android, and desktop without platform-specific SDKs.
    
    \item \textit{Zero App Store Friction}: Instant installation via single tap, no download waits, no app store review delays for updates.
\end{itemize}

\subsubsection{Trade-offs}

\begin{itemize}
    \item \textit{iOS PWA Limitations}: iOS restricts some PWA capabilities compared to Android (e.g., limited storage quotas, notification restrictions). However, iOS is improving PWA support incrementally.
    
    \item \textit{Browser Engine Dependencies}: PWAs depend on browser capabilities. Safari (iOS's only browser engine) occasionally lags Chrome in feature support.
\end{itemize}

\subsubsection{Why It's Worth It}

\begin{itemize}
    \item \textit{Distribution Flexibility}: PWAs install without app stores, work in embedded contexts (Telegram, TikTok), and update instantly.
    
    \item \textit{Development Efficiency}: One codebase instead of three (iOS, Android, web). Faster iteration cycles.
    
    \item \textit{User Friction Reduction}: No 100-200MB app downloads, no app store navigation, no update delays.
    
    \item \textit{Mobile-Specific Optimization}: PWAs are ideal for mobile Web3 specifically. They work in super app ecosystems (Telegram, WeChat), install instantly (critical for Web3's already-high CAC), and maintain always-updated code (important for security patches).
\end{itemize}

\subsection{Parent Application Dependency}

dApps depend on Refract Passport for wallet operations, transaction signing, and platform API access.

\subsubsection{Rationale}

\begin{itemize}
    \item \textit{Security Isolation Requirement}: Wallet operations must occur outside dApp control to prevent manipulation. The parent-child architecture is the only mobile-compatible model providing this isolation.
    
    \item \textit{Shared Wallet Across dApps}: Users have one wallet across all dApps, unlike embedded wallets' per-dApp isolation. This requires a shared parent application managing the wallet.
    
    \item \textit{Native Platform API Brokering}: Push notifications, home screen installation, and other native APIs must be brokered by the parent application due to browser security policies.
    
    \item \textit{MPC Key Share Storage}: Key shares are stored in the parent application's secure storage, inaccessible to individual dApps.
\end{itemize}

\subsubsection{Graceful Degradation}

If a user opens a Refract dApp's URL directly in a standard browser (outside Passport), the Refract SDK detects absence of the parent context and does not inject its provider. Standard browser extension wallets (MetaMask, Phantom, Coinbase Wallet) remain discoverable via EIP-6963 \cite{eip6963}. The dApp functions normally with these wallets, just without SecureSign-specific features. This ensures dApps remain functional on desktop and for users preferring traditional wallets.

\subsubsection{Why This Is Acceptable}

\begin{itemize}
    \item \textit{Desktop Browser Extensions Create Identical Dependency}: MetaMask, Phantom, and other browser extensions create the same architecture—webpages depend on the extension for wallet operations. This proven model achieves strong security.
    
    \item \textit{Refract Replicates Proven Model on Mobile}: Browser extensions don't exist on mobile. Refract provides the mobile equivalent.
    
    \item \textit{Users Choose Refract for Its Benefits}: Users install Refract Passport for superior mobile UX. Desktop fallback is secondary.
\end{itemize}

\subsection{Why Limitations Are Actually Strengths}

These architectural requirements enable SecureSign's core value propositions:

\begin{itemize}
    \item \textit{Magic dApp Links Enable}:
    \begin{itemize}
        \item Verified application identity preventing phishing
        \item PWA installation providing home screen shortcuts
        \item Deep linking enabling notifications and social sharing
        \item Sandboxed security isolating dApps from each other
    \end{itemize}
    
    \item \textit{PWA Architecture Enables}:
    \begin{itemize}
        \item Zero app store friction with instant installation
        \item Native platform APIs for notifications and sharing
        \item Cross-platform consistency with one codebase
        \item Embedded distribution inside super apps
    \end{itemize}
    
    \item \textit{Parent Application Dependency Enables}:
    \begin{itemize}
        \item Desktop-class security with sandboxed signing context
        \item Shared wallet with interoperable assets across dApps
        \item Native capabilities through platform API brokering
        \item Graceful fallback working with browser extensions
    \end{itemize}
\end{itemize}

These are \textit{mobile-first design choices} optimizing for mobile device constraints and opportunities. Desktop Web3 has browser extensions; mobile Web3 needs a different model. SecureSign provides that model without compromising security or developer experience.

\section{Discussion}

\subsection{Ecosystem Implications}

SecureSign's architecture has broader implications for the mobile Web3 ecosystem beyond individual dApp user experiences.

\begin{itemize}
    \item \textit{Distribution Channel Emergence}: Refract Passport becomes a discovery platform for mobile Web3 dApps, similar to how app stores aggregate mobile apps. This provides distribution channels currently absent for mobile dApps (Section 3.1), reducing customer acquisition costs.
    
    \item \textit{Shared Infrastructure Benefits}: All dApps using SecureSign share authentication infrastructure, wallet management, and security auditing. Improvements to Passport benefit the entire ecosystem, creating network effects.
    
    \item \textit{Developer Ecosystem}: Standardizing on EIP-6963 and PWA architectures creates transferable knowledge. Developers skilled in building Refract dApps can easily build additional dApps, growing the ecosystem.
    
    \item \textit{Interoperability by Default}: Unlike embedded wallets where each dApp has isolated wallets, SecureSign provides one wallet across all dApps. Users carry assets, identity, and reputation seamlessly, enabling composability and network effects central to Web3 principles.
\end{itemize}

\subsection{Developer Adoption Considerations}

SecureSign's success depends on developer adoption. Several factors favor adoption:

\begin{itemize}
    \item \textit{Zero Migration Cost}: Existing Web3 dApps integrate SecureSign with 5-10 lines of code wrapping existing providers. No codebase overhaul required (contrast: embedded wallets requiring complete architecture migration).
    
    \item \textit{Maintained Compatibility}: dApps continue working with browser extension wallets on desktop via EIP-6963 fallback. Developers add mobile support without losing desktop functionality.
    
    \item \textit{Improved Retention = Better Economics}: By solving retention barriers (push notifications, zero context switching, home screen presence), SecureSign improves dApp unit economics. Higher lifetime value justifies customer acquisition costs.
    
    \item \textit{No Vendor Lock-In for Users}: While dApps depend on Refract Passport, \textit{users are not locked in}. Wallet key shares can be exported, and users can migrate to other wallet solutions if desired. This reduces adoption risk for developers concerned about centralized dependencies.
\end{itemize}

\subsection{Limitations and Challenges}

Despite its advantages, SecureSign faces limitations:

\begin{itemize}
    \item \textit{iOS PWA Restrictions}: Apple's App Store policies and iOS Safari limitations constrain PWA capabilities. While improving, iOS PWAs have reduced storage quotas, more limited notification support, and occasional feature regressions when iOS updates. SecureSign's iOS experience is thus somewhat degraded compared to Android.
    
    \item \textit{Centralized Infrastructure Dependency}: Refract's backend infrastructure brokers MPC key shares, handles push notifications, and maintains the dApp registry. While this provides better UX than fully decentralized alternatives, it introduces centralized failure points. Infrastructure outages, regulatory challenges, or business continuity issues could impact all SecureSign dApps.
    
    \item \textit{New Mental Model for Users}: Users must learn that dApps are accessed through Refract Passport rather than directly via URLs or app store downloads. This adds initial education overhead, though Passport's UX improvements quickly provide value justifying the learning curve.
    
    \item \textit{Market Competition with Established Wallets}: MetaMask Mobile, Phantom, and Rainbow have existing user bases and brand recognition. Convincing users to switch to Refract Passport requires demonstrating clear value superiority—addressed through superior retention mechanisms (Section 6).
\end{itemize}

\section{Future Work}

SecureSign's current implementation provides a foundation for additional capabilities:

\subsection{Enhanced EIP-6963 Integration}

Full implementation of EIP-6963's multi-provider discovery would allow SecureSign to coexist with other mobile wallet solutions, letting users choose their preferred wallet while still accessing dApps through Refract Passport. This would reduce switching costs for users with existing wallet preferences.

\subsection{Deep PWA Features}

Further leveraging PWA capabilities:
\begin{itemize}
    \item \textit{Background Sync}: Queue transactions when offline, auto-submit when connectivity restored
    \item \textit{Persistent Storage}: Larger storage quotas for data-intensive dApps (gaming, social)
    \item \textit{Geolocation and Sensors}: Enable location-based dApps and AR/VR Web3 experiences
\end{itemize}

\subsection{Social Identity Integration}

Integrate decentralized identity protocols (ENS, Lens Protocol, Farcaster) allowing users to carry social identities across dApps. Users could log in with social handles rather than wallet addresses, improving familiarity for Web2 users.

\subsection{Developer Tooling}

\begin{itemize}
    \item \textit{Setup Wizard}: Interactive CLI tool guiding developers through SecureSign integration, generating boilerplate code, and handling dApp registration.
    
    \item \textit{Development Sandbox}: Local testing environment simulating Refract Passport's iframe sandbox and \texttt{postMessage} communication, enabling developers to test SecureSign integration without deploying.
    
    \item \textit{Analytics Dashboard}: Provide developers with retention analytics, funnel visualization, and notification performance metrics, helping optimize dApp engagement.
\end{itemize}

\subsection{Ecosystem Expansion}

\begin{itemize}
    \item \textit{Cross-Chain Support}: Extend beyond Ethereum to support Solana, Polygon, Arbitrum, and other chains with unified multi-chain wallet management.
    
    \item \textit{Fiat On-Ramp Integration}: Directly integrate fiat-to-crypto on-ramps within Passport, allowing users to purchase crypto without leaving the dApp experience.
    
    \item \textit{dApp Store}: Build curated discovery platform within Passport showcasing high-quality dApps, providing the distribution channel currently absent for mobile Web3.
\end{itemize}

\section{Conclusion}

Mobile Web3 faces a catastrophic retention crisis rooted in fundamental architectural incompatibility between mobile platform constraints and Web3 security requirements. Existing solutions force an impossible tradeoff: embedded wallets sacrifice security through unfixable iframe vulnerabilities to achieve moderate usability, while app wallets maintain security but suffer from 2-3\% retention at 30 days due to download friction and context-switching penalties.

We presented SecureSign, a novel mobile Web3 architecture that resolves this security-UX incompatibility by adapting desktop browser extension security models to mobile through Progressive Web App based sandboxing. By isolating dApp execution in iframes within a parent Refract Passport application, we achieve desktop-class security properties—transaction approval integrity, click-jacking immunity, credential protection—while providing superior mobile UX through native platform capabilities including push notifications, home screen installation, zero context switching, native share APIs, and offline caching.

SecureSign provides drop-in integration requiring zero codebase changes for existing Web3 applications, eliminates the costly migration penalties embedded wallets impose, maintains wallet interoperability across dApps consistent with Web3 principles, and directly addresses retention barriers responsible for mobile Web3's 3.55\% six-month retention rate.

Our threat model analysis demonstrates immunity to click-jacking, transaction overlay, credential skimming, and API impersonation attacks while maintaining graceful fallback to browser extension wallets on desktop. SecureSign represents a fundamental architectural innovation enabling mobile Web3 to achieve the security guarantees and user experiences necessary for mainstream adoption.

The mobile Web3 opportunity is massive—over 50\% of internet traffic, billions of potential users, high willingness to pay for quality experiences. SecureSign provides the architectural foundation to capture this market by finally resolving the security-UX incompatibility that has prevented mobile Web3 from reaching its potential.

%%%%%%%%%%%%%%%%%%%%%%%%%%%%%%%%%%%%%%%%%%%%%%%%%%%%%%%%%%%%%%%%%%%%%%%%%%%%%%%%
% BIBLIOGRAPHY
%%%%%%%%%%%%%%%%%%%%%%%%%%%%%%%%%%%%%%%%%%%%%%%%%%%%%%%%%%%%%%%%%%%%%%%%%%%%%%%%

% This tells LaTeX to use the IEEEtran bibliography style
\bibliographystyle{IEEEtran}

\bibliography{refs}

\end{document}